# Encapsulation epitaxy of air-stable monolayer superconducting films for quantum circuits and qubits

Xudong Zheng[1,‡], Sameia Zaman[1,2,‡], Kenan Zhang[1,‡], Connor A Occhialini[3], Haowei Xu[4], Zhien Wang[5], Fangyuan Liu[6], Luiz Gustavo Pimenta Martins[7], Sejoon Lim[7], Tianyi Zhang[1], Tilo H. Yang[1], Jiangtao Wang[1], Yunyue Zhu[1], Zachariah Hennighausen[1], Sein Park[2,8], Steven Vitale[9], Kevin Tibbetts[9], Stephen Margiotta[9], Phillip Kim[7], Cong Su[6], Ju Li[4], Riccardo Comin[3], William D. Oliver[1,2,3*], Joel Î-j. Wang[2,10,*], Jing Kong[1,2,*]

[1] Department of Electrical Engineering and Computer Science, Massachusetts Institute of Technology, Cambridge, MA, USA

[2] Research Laboratory of Electronics, Massachusetts Institute of Technology, Cambridge, MA, USA

[3] Department of Physics, Massachusetts Institute of Technology, Cambridge, MA, USA

[4] Department of Nuclear Science and Engineering, Massachusetts Institute of Technology, Cambridge, MA, USA

[5] Department of Materials Science and Engineering, Massachusetts Institute of Technology, Cambridge, MA, USA

[6] Department of Material Science, Yale University, New Haven, CT, USA

[7] Department of Physics, Harvard University, Cambridge, MA, USA

[8] Department of Physics, Pohang University of Science and Technology, Pohang, Republic of Korea

[9] Lincoln Laboratory, Massachusetts Institute of Technology, Lexington, MA, USA

[10] Department of Physics, New York University, New York, NY, USA

[*] Corresponding authors: wi18222@mit.edu (WO); joelwang@mit.edu (JW); jingkong@mit.edu (JK)

[‡] These authors contributed equally to this work.

## Abstract

Two-dimensional (2D) superconductors are an emerging platform supporting both strongly correlated physics and quantum information science. Their reduced dimensionality, atomically flat interfaces, and high-crystallinity are particularly attractive for realizing compact lumped-element devices in superconducting circuits. However, successful synthesis of large-area, monolayer 2D superconductors remains challenging as they are easily oxidized in air. In this work, we found a novel "encapsulation epitaxy" mechanism that enables the growth of large-area, air-stable, monolayer superconducting $NbSe_2$ films and explored their potential use in fabricating superconducting quantum circuits. This work represents a completely new growth phenomenon in which a 2D encapsulation layer, such as graphene or hexagonal boron nitride (hBN), pre-deposited on a 3D substrate (e.g., $SiO_2$, $Si_3N_4$) can act both as a template for the epitaxial growth of monolayer $NbSe_2$ (1L-$NbSe_2$) underneath it (at the encapsulation-substrate interface) and as a protection cover on top. This approach enables the formation of large area (>1-inch) uniform 1L-$NbSe_2$ with greatly enhanced ambient stability and thereby device fabrication in air, owing to the inherent protection provided by the pre-formed encapsulating layer. Using such grown 1L-graphene/$NbSe_2$ heterostructures, we observe robust superconductivity (superconducting transition temperature $T_c \approx 1$ K) and enhanced charge density wave (CDW; CDW transition temperature $T_{CDW} \approx 177$ K), which are sensitive to disorders and are indicative of well-structured material. We further demonstrate the integration of 1L-$NbSe_2$ into superconducting circuits by developing oxidation-free transfer and superconducting edge-contact techniques. The 1L-$NbSe_2$ in these circuits feature a measured kinetic inductance $L_K \approx 0.7$ nH/□, making this material suitable for quantum circuits requiring circuit elements with high kinetic inductance. This encapsulation epitaxy methodology enables the production of air-stable 2D superconductors and van der Waals heterostructures, holding the promise for wafer-scale, monolithic fabrication of superconducting quantum circuitry.

# Main text

2D superconducting transition metal dichalcogenides (SC-TMDs) are materials system where strongly correlated physics and quantum information science converge[1,2]. At reduced dimensionality, quantum confinement significantly modifies the behavior of electronic phases in SC-TMDs, such as charge density waves[3], quantum spin liquids[4], and superconductivity[5], leading to distinct phenomena compared to their bulk counterparts, making 2D SC-TMDs an ideal platform for studying correlated physics. The atomically flat interfaces and layered structure of SC-TMD heterostructures make them promising candidates for realizing quantum devices with significantly reduced footprint and crosstalk[6–8]. For instance, superconducting transmon qubits incorporating $NbSe_2$/hBN/$NbSe_2$ parallel-plate capacitors have achieved coherence times comparable to those of conventional aluminum-based qubits, while occupying over 250 times less area[6]. However, monolayer SC-TMDs are highly air sensitive, and their superconductivity is easily disrupted by external perturbations and local defects. The growth of high-quality, monolayer SC-TMDs has been a critical challenge[9] hindering their application toward next generation quantum devices.

Unlike semiconducting TMDs (such as $MoS_2$ and $WSe_2$), which have seen rapid progress in synthesis techniques[10–12], the difficulties in the growth of monolayer SC-TMDs come from the high melting points and low surface diffusivity of the precursors, which hinder the formation of continuous and uniform monolayers[13]. For instance, attempts to synthesize monolayer SC-TMDs using molecular beam epitaxy (MBE)[14,15] or molten-salt assisted chemical vapor deposition (CVD)[13,16] have only yielded micrometer-sized flakes with non-uniform layer thickness. While electrochemical exfoliation produces monolayer flakes with larger lateral sizes, the thickness and uniformity of these films are poorly controlled in general[17]. In another two-step CVD approach, a metal film is first deposited on the target substrate and subsequently converted into SC-TMDs, which has enabled the synthesis of wafer-scale SC-TMDs[18,19]. However, the average grain sizes of these films are in the nanometer regime (≈ 8.2 nm for monolayer $NbSe_2$[19] and ≈ 50-100 nm for multilayers SC-TMDs[18]), which compromises the material quality (e.g. more scattering at grain boundaries). To date, there is no reliable way to grow homogeneous, highly-crystalline, and monolayer SC-TMDs films at the wafer-scale.

In addition to superconductivity, SC-TMDs usually exhibit CDW order[20]. A CDW is the periodic modulation of the conduction electron density and its associated lattice distortion. At atomically thin thickness, the electron-phonon interaction is strengthened and leads to more prominent CDW phenomena. For instance, 1L-$NbSe_2$ has been reported to exhibit strongly enhanced CDWs with transition temperatures ($T_{CDW}$) as high as 145 K[3]. However, as the CDW phase transition requires materials with low disorder[20,21], the observation of such phenomena has been limited primarily to exfoliated and MBE-grown samples[3,14,15]. CVD-grown materials, in comparison, tend to have relatively higher levels of disorder, which may disrupt the long-range coherence that is critical for CDWs formation[21].

In this work, we report an "encapsulation epitaxy" methodology that enables the growth of large-area monolayer $NbSe_2$ film at the interface of a hybrid 2D/3D substrate, comprising a 2D layer (graphene or hBN; serving both as growth template and subsequently as encapsulation layer) and a 3D amorphous $SiO_2$ substrate. Density functional theory (DFT) calculations show the 2D/3D substrate combine the advantages of a low diffusion barrier present at 2D material surfaces and a high adsorption energy present at 3D $SiO_2$ surfaces, which together ensures monolayer growth in the van der Waals gap with ultra-clean interfaces. A combination of various optical, electrical, and electron microscopy tools confirmed the exceptional material quality, as discussed below, of the as-synthesized monolayer SC-TMDs. We observed robust CDW formation comparable to exfoliated samples in both CVD graphene/$NbSe_2$ and hBN/$NbSe_2$. We further demonstrate a reliable process for integrating our CVD-grown 1L-$NbSe_2$ into superconducting circuits, enabled by oxidation-free, wafer-scale transfer of SC-TMD thin films and by a superconducting, one-dimensional edge-contact technique that exhibits negligible contact resistance at low temperatures. Finally, we extract the kinetic inductance of the CVD monolayer $NbSe_2$ via microwave characterization, highlighting the potential of CVD-grown SC-TMDs as a material platform for developing extensive quantum computing architectures based on two-dimensional materials.

**Encapsulation epitaxy**

The "encapsulation epitaxy" growth process is illustrated in Fig. 1a. First, a layer of 2D material (such as graphene or hBN) is transferred to a 3D amorphous $SiO_2$ substrate. Then this 2D-3D hybrid substrate is placed inside a CVD furnace for the synthesis of 1L-SC. Under the growth conditions we have designed for this work (see Methods for details), no growth happens on the exposed 2D surface. Instead, the precursors enter the van der Waals gap at 2D-3D interface, nucleate and form epitaxial 1L-SC flakes, and eventually merging into a continuous 1L-SC film at the interface. No multilayer growth is observed during this process (Fig. 1b). Such growth behavior is not observed on the exposed 3D or 2D surfaces under the same growth conditions. For example, on the exposed 3D surface, only isolated thick flakes with non-uniform thickness, typically >100 nm, is detected (see Extended Data Fig. 1; note this growth was performed under a metal-precursor-rich condition, see Methods section "selective growth" for further discussions). On the exposed 2D surface, no growth of 1L-SC is observed, as evidenced by Raman characterization (see Extended Data Fig. 2).

These growth behaviors are elucidated by DFT calculations in combination with thermodynamic and kinetic modeling[22,23]. In CVD thin film growth, the surface diffusion coefficient $D_s$ determines how quickly adatoms can travel on the surface and follows an Arrhenius form[24]:

$$D_s = D_0 \cdot e^{-E_{\text{diff}}/k_B T} \quad (1)$$

where $D_0$ is the pre-exponential factor, $E_{diff}$ is the surface diffusion energy barrier, $k_B$ is the Boltzmann constant, and $T$ is the growth temperature. Meanwhile, the residence time $\tau$ determines how long adatoms can stay on the surface before desorbing (thus the likelihood of adatom incorporation) and is defined[23]:

$$\tau = \nu^{-1} \cdot e^{E_{\text{ads}}/k_B T} \quad (2)$$

where $\nu$ is the surface vibration frequency and $E_{ads}$ is the adsorption (binding) energy between the adatom and the substrate. These equations connect the surface energies ($E_{diff}$, $E_{ads}$) with the kinetic parameters ($D_s$, $\tau$).

On the exposed $SiO_2$ surface, DFT calculations reveal a high diffusion barrier for $NbSe_2$ adatoms ($E_{diff} \approx 1.0$ eV), which greatly exceeds the available thermal energy at the relevant growth temperature (950 °C). According to equation (1), the diffusion coefficient is relatively low ($D_s \approx 7.6 \times 10^{-8}$ cm$^2$/s). It severely limits the lateral adatom migration that is critical for layer-by-layer growth. Meanwhile, $NbSe_2$ binds strongly to $SiO_2$ ($E_{ads} \approx 2.6$ eV), resulting in a long residence time ($\tau \approx 5.2$ ms) that promotes nucleation. These make vertical growth more favorable than lateral growth, leading to the formation of thick flakes on bare $SiO_2$. In contrast, the graphene surface presents a lower diffusion barrier ($E_{diff} \approx 0.3$ eV), enabling fast surface diffusion with a diffusion coefficient over three orders-of-magnitude higher than on $SiO_2$ ($D_s \approx 5.8 \times 10^{-5}$ cm$^2$/s). However, graphene's weaker adsorption energy ($E_{ads} \approx 1.9$ eV) results in a three-orders-of-magnitude shorter residence time ($\tau \approx 6.8 \times 10^{-3}$ ms). This increases the desorption rate and significantly suppresses nucleation, resulting in minimal growth on graphene. In short, $SiO_2$ favors nucleation but hinders lateral growth; graphene supports adatom mobility but fails to sustain nucleation. Encapsulation epitaxy overcomes these limitations via a 2D–3D hybrid substrate design, which combines the high adsorption energy of the 3D substrate surface to promote a sufficient residence time and nucleation, and a low diffusion barrier of 2D layer surface for efficient lateral migration. This synergistic effect enables uniform monolayer formation at the 2D–3D interface (additional DFT calculations and substrate dependent growth results shown in Extended Data Fig. 3&4; additional discussions provided in Methods).

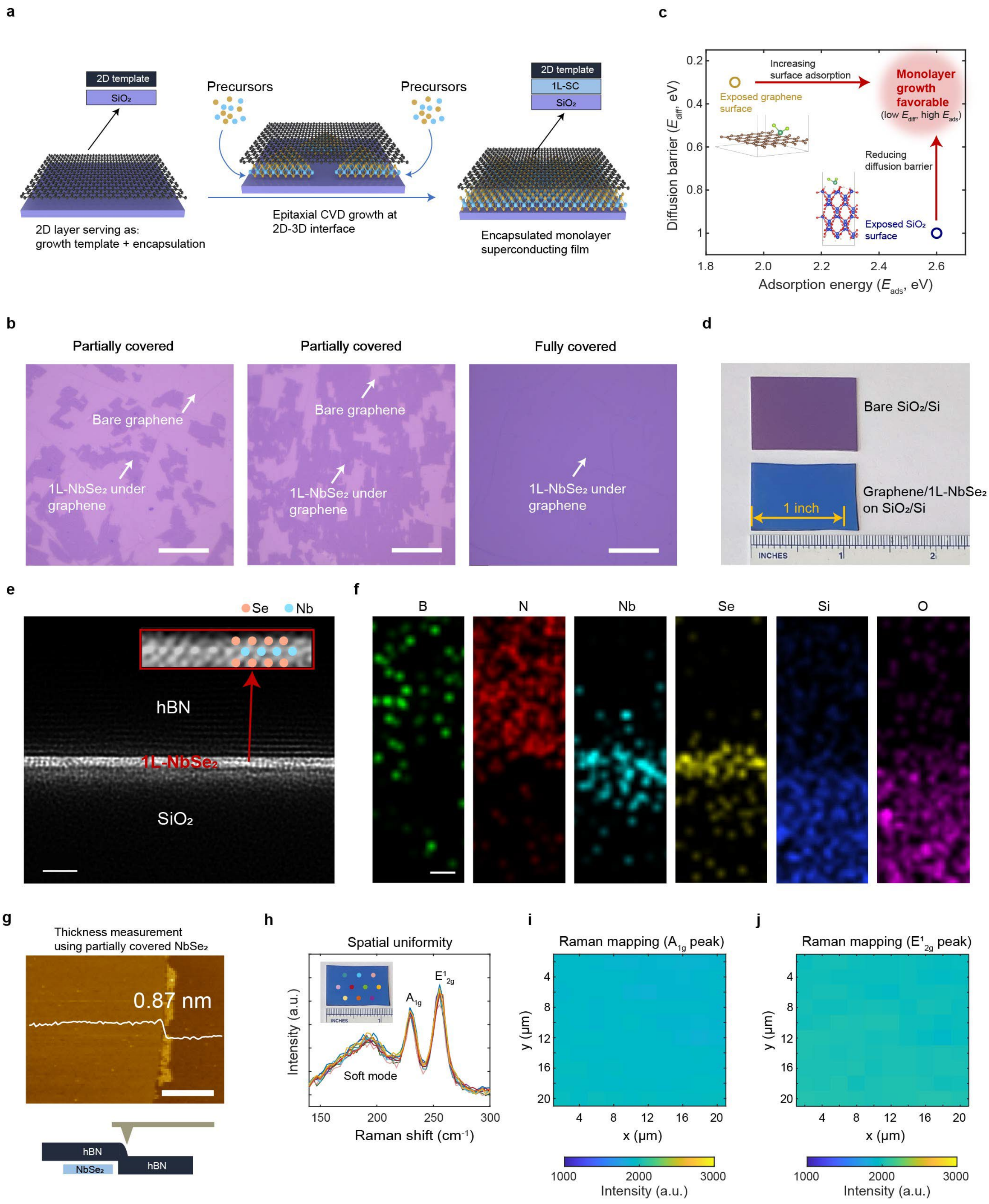

**Fig. 1 | Encapsulation epitaxy: growth process and resulting 1L-$NbSe_2$ film. a,** Schematic illustration of the encapsulation epitaxy growth process. The growth of monolayer SC happens in the van der Waals gap between

the 2D template layer and the 3D amorphous $SiO_2$ substrate. **b,** Optical micrographs of CVD-grown 1L-$NbSe_2$ with different growth coverage, confirming the growth of 1L-$NbSe_2$ only with no multilayer growth. Additional characterization of the partially grown $NbSe_2$ can be found in Supplementary Fig. 1 and in Methods. Scale bar, 20 μm. **c,** DFT-calculated surface diffusion barrier and adsorption energy of $NbSe_2$ on exposed graphene (yellow circle) and $SiO_2$ (blue circle) surfaces. The red-shaded "Monolayer growth favorable" region is a qualitative guide, not derived from DFT. **d,** Photo showing the as-grown graphene/$NbSe_2$/$SiO_2$/Si and its comparison with bare $SiO_2$/Si. **e,** Cross-sectional scanning transmission electron microscope (STEM) image of 1L-$NbSe_2$ grown at the interface between an hBN encapsulation layer and a $SiO_2$ substrate. The region on the top of the 1L-$NbSe_2$ in the image is fully occupied by hBN, but the contrast is low as it is adjusted for optimal visualization of $NbSe_2$. Scale bar, 2 nm. Inset, a zoomed-in image of the 1L-$NbSe_2$ cross section grown underneath hBN. **f,** EDS mappings for the sample in panel **1e**. Scale bar, 2 nm. **g**, AFM measurement of a 1L-$NbSe_2$ grown underneath hBN via encapsulation epitaxy. A partially covered $NbSe_2$ sample is used, i.e., regions with only hBN and regions with hBN/$NbSe_2$, to enable the measurement of the 1L-$NbSe_2$ thickness from the height difference of the hBN surface as illustrated in the bottom diagram. Scale bar: 500 nm. **h,** Spatial uniformity of $NbSe_2$ characterized by Raman spectroscopy at 10 different locations across the sample shown in panel **1d**. The measurement locations are indicated by colored dots in the inset photo. **i-j,** Raman mapping of 20×20 $\mu m^2$ area indicating the uniformity of the $NbSe_2$ (optical image and all Raman spectra shown in Extended Data Fig. 5c-d).

Figure 1d shows the photograph of inch-scale CVD-grown 1L-$NbSe_2$ at the graphene/$SiO_2$ interface. The growth is enabled by innovations in both substrate engineering and precursor delivery. The substrate employs the special 2D-3D hybrid approach, as discussed in the previous section. The precursor delivery is based on a modified face-to-face strategy and utilizes vapor-phase salt as growth promoters (see Supplementary Fig. 2 and Methods), enabling highly scalable and uniform growth. The as-synthesized $NbSe_2$ films exhibit excellent growth uniformity (Fig. 1h-j and Extended Data Fig. 5c-d) and monolayer thickness, as determined by cross-sectional high-angle annular dark-field scanning transmission electron microscopy (HAADF-STEM) imaging (Fig. 1e), atomic force microscopy (AFM) measurements (Fig. 1g), and the absence of an interlayer shear mode between 20–30 $cm^{-1}$ in the Raman spectra (Fig. 2c). The material also shows well-defined chemical structure, supported by Raman and X-ray photoelectron spectroscopy (XPS) analyses (Extended Fig. 5). The fact that growth occurs at the 2D–3D interface, rather than on the exposed top surface of the 2D template, is confirmed by a combination of cross-sectional STEM imaging (Fig. 1e) and the corresponding energy-dispersive X-ray spectroscopy (EDS) mapping (Fig. 1f), transfer test (Extended Data Fig. 2) and air-stability characterization (Extended Data Fig. 6). Notably, the encapsulation layer significantly enhances the air-stability of $NbSe_2$ throughout the growth and post-growth characterization and fabrication processing, including robust Raman characterization in ambient even after the sample was kept in air for 2 weeks (Extended Data Fig. 6). When we peeled off the encapsulation layer, which can be done for $NbSe_2$ grown at the hBN-$SiO_2$ interface (Extended Data Fig. 2), only the hBN was picked up while $NbSe_2$ stayed on the $SiO_2$. After hBN removal, the exposed 1L-$NbSe_2$ starts to degrade immediately due to contact with the air. Ambient Raman characterization becomes no longer feasible, because even low-power and short-time laser excitations (≈140 μW/μm$^2$ for 3 s) will create a hole in the un-encapsulated $NbSe_2$ (Extended Data Fig. 2); this is one example that clearly illustrates the ambient stability enabled by encapsulation epitaxy.

In Fig. 2a, the HAADF STEM image presents the top-view, atomic-scale lattice structure of as-synthesized 1L-$NbSe_2$, with a well-defined hexagonal lattice for 1L-$NbSe_2$ with 1H crystal structure. The extracted lattice constant is 0.34 nm, in good agreement with prior literature[25]. The encapsulation graphene is not visible in the HAADF-STEM image due to the much smaller atomic number of carbon as compared to niobium and selenium. Fig. 2b shows the selected area electron diffraction (SAED) pattern of $NbSe_2$, which clearly shows the diffraction pattern of both the $NbSe_2$ and encapsulation graphene layers. Additional SAED images are provided in Extended Data Fig. 7. These diffraction patterns are taken across a 200 × 200 μm$^2$ area, and the aligned patterns of $NbSe_2$ and graphene indicate the $NbSe_2$ is epitaxially aligned with the graphene lattice. This is consistent with the recent "hypotaxy" growth of single-crystalline $MoS_2$ underneath graphene[26]. However, unlike the previous approach where metal precursors were pre-deposited on the substrate prior to graphene encapsulation, in our approach the precursors are delivered in gas phase directly into the pre-formed graphene/$SiO_2$ or hBN/$SiO_2$ interface, which facilitates monolayer growth. In addition, the top graphene (or hBN) layer remains after growth is finished, also serving as an effective encapsulation layer afterwards. This approach further supports the selective-area growth of $NbSe_2$ using patterned graphene on a $SiO_2$ substrate (see Extended Data Fig. 8; see also Methods for details of controlling the gas-phase metal precursor concentration in the low-mass-flux regime), which indicates the possibility of obtaining patterned structures directly from growth to facilitate specific fabrication needs in the future. Moreover, this approach can also be extended to the growth of other high-melting-point TMD materials (see Extended Data Fig. 9 for preliminary results).

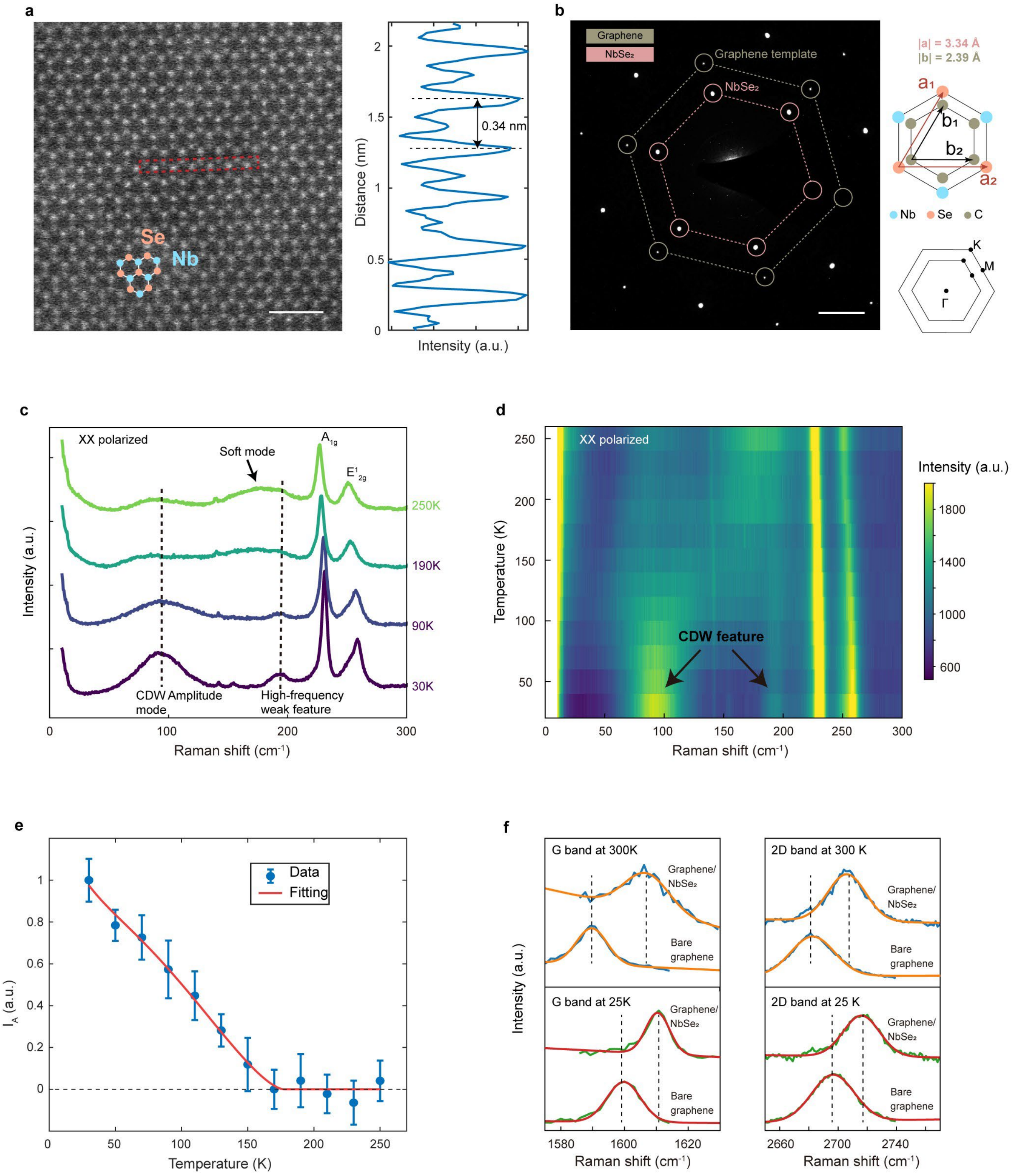

**Fig. 2 | Atomic imaging and charge density wave characterization. a,** (left panel) HAADF-STEM image of a 1L-graphene/$NbSe_2$ sample showing the lattice structure of $NbSe_2$. The graphene is not visible due to the much lower atomic number of carbon compared to niobium and selenium. Scale bar, 1 nm. (right panel), Line profile from the region outlined by the red box in the STEM image. **b**, Selective area electron diffraction (SAED) pattern of the 1L-graphene/$NbSe_2$. Scale bar, 2 $nm^{-1}$ **c,** Temperature-dependent high-resolution Raman spectra on 1L-graphene/$NbSe_2$ heterostructure synthesized by encapsulation epitaxy. The symbol XX denotes that the incident excitation laser is linearly polarized, and only the linear, parallel polarization is collected at the spectrometer entrance slit. **d,** Temperature map of the Raman scattering intensity of 1L-graphene/$NbSe_2$ heterostructure. **e,**

Temperature dependence of the CDW amplitude mode intensity ($I_A$) for a 1L-graphene/$NbSe_2$ heterostructure (see Methods for extraction details). The error bars are calculated based on the standard deviation of the Raman scattering intensity around the amplitude mode peak and the spectral width employed for integrating the mode. Solid lines are fits to mean field theory[3] and serve as guides to the eye. **f,** Comparison between 1L-graphene/$NbSe_2$ and bare 1L-graphene in terms of the graphene Raman vibrational modes at both room temperature (300 K) and cryogenic temperature (25 K). Gaussian fits with a linear background are overlaid to locate the peak positions.

**Charge density wave in CVD 1L-$NbSe_2$ films**

The CVD-grown 1L-$NbSe_2$ by encapsulation epitaxy exhibits exceptional crystalline quality, as manifested by the robust CDW signatures comparable to, or even surpassing, those of exfoliated 1L-$NbSe_2$. Fig. 2c-d show CDW measurements of 1L-graphene/$NbSe_2$ on $SiO_2$/Si substrate via temperature-dependent Raman characterization from 30 K to 250 K (additional characterization shown in Supplementary Fig. 3). The monolayer thickness of $NbSe_2$ is validated by the absence of an interlayer shear mode[3] between 20–30 $cm^{-1}$. As temperature decreases, we observe the clear emergence of a CDW amplitude mode and a weak, high-frequency feature (marked in the figure), two characteristics of CDWs in $NbSe_2$. The extracted transition temperature ($T_{CDW}$) for graphene/$NbSe_2$ is ≈ 177.0 K (Fig. 2e). Similar results are observed on 1L-$NbSe_2$ grown at the hBN/$SiO_2$ interface (Extended Data Fig. 10, $T_{CDW}$ ≈ 185.1 K). These features are in good agreement with prior literature measuring exfoliated 1L-$NbSe_2$[3]. Since CDWs are readily weakened and destroyed by disorders[21], the observation of a strong CDW signal indicates the high crystalline quality of our CVD grown 1L-$NbSe_2$ with low defect density. Additionally, since the graphene/$NbSe_2$ and hBN/$NbSe_2$ heterostructures are formed directly through epitaxial growth, the resulting interfaces are atomically clean and free of bubbles—an essential requirement for the study of strongly correlated phenomena. Both the G and 2D bands of the graphene on top of $NbSe_2$ exhibit pronounced blue shifts compared to graphene on $SiO_2$ (Fig. 2f and Supplementary Fig. 4), indicating high tensile strain levels of graphene (greater than approximately 0.6%)[27,28], and is attributed to larger lattice constant of the underlying $NbSe_2$.

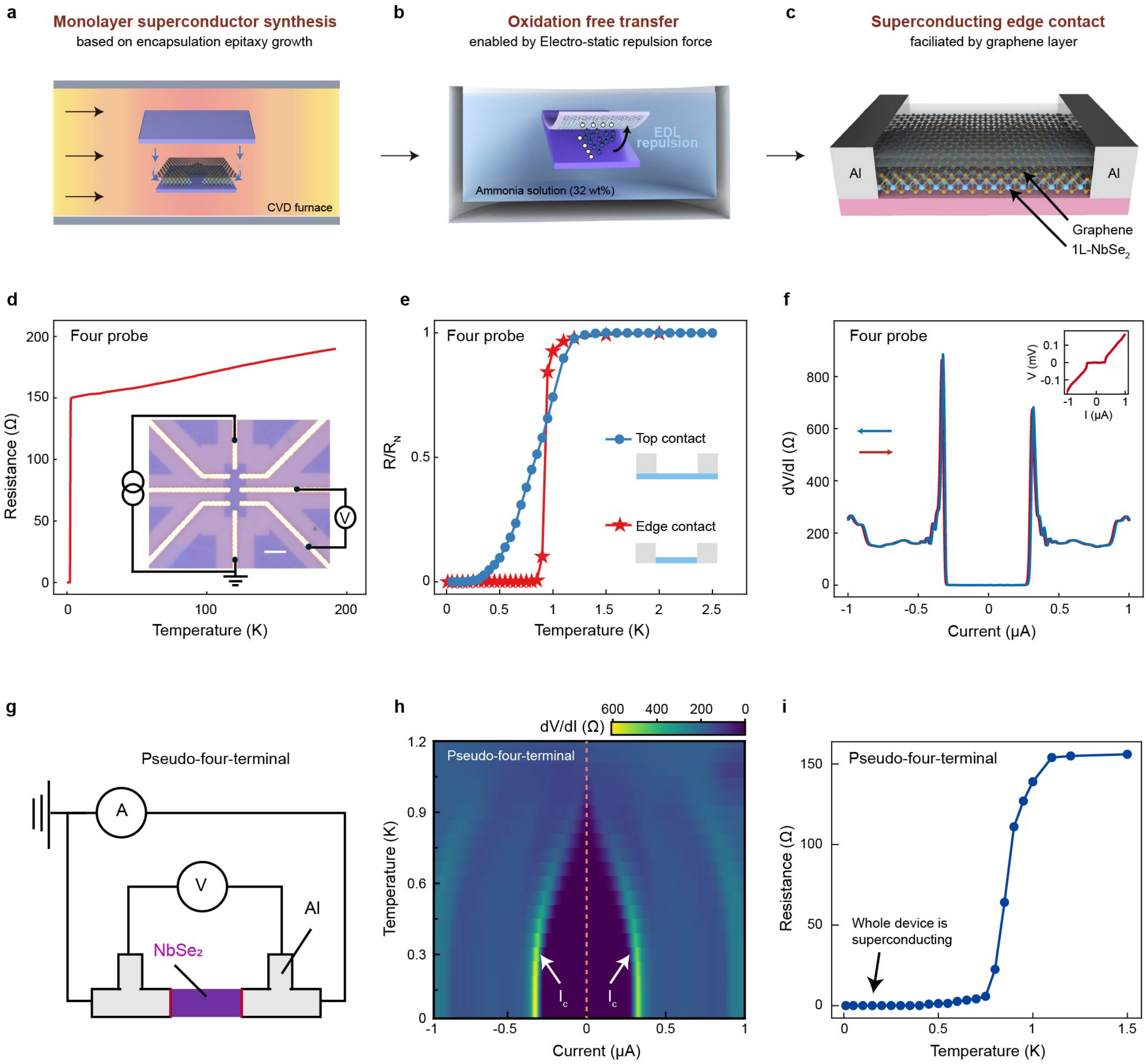

**Fig. 3 | Superconductivity and superconducting edge contact. a-c,** Schematics of the fabrication process developed in this work for integration into superconducting circuits, which includes large-area synthesis of 1L-$NbSe_2$ using encapsulation epitaxy, oxidation-free transfer of 1L-$NbSe_2$ to target substrates, and graphene-facilitated superconducting edge contact. **d,** Four-probe resistance vs. temperature measurements of the 1L-graphene/$NbSe_2$ film with edge contacts. Inset, optical image of the fabricated device. Scale bar, 5 µm. **e,** Superconducting transition of 1L-$NbSe_2$ with two different contact configurations: the edge-contacted device (red trace) exhibits a sharp transition while the top-contacted device (blue trace) shows a softer transition to zero resistance. **f,** Differential resistance (d$V$/d$I$) as a function of DC-bias current measured at 10 mK using an AC excitation of 5 nA. The data was acquired for both increasing (red) and decreasing (blue) current sweeps. Inset: Current-voltage ($I$-$V$) curves measured in both sweeping directions exhibit no discernible hysteresis within the resolution of this measurement. **g,** Schematic of the pseudo-four-probe configuration used to characterize the superconducting edge-contact to Al electrodes**. h,** Differential resistance (d$V$/d$I$) measured in a pseudo-four-probe configuration as a function of temperature from 150 mK to 1.2K. The d$V$/d$I$ approaches zero in the dark blue region, indicating zero series resistance at the Al and $NbSe_2$ edge-contacted interfaces. **i,** Temperature dependence of the differential resistance at zero DC bias ($I_{DC}$ = 0 µA line-cut in Fig. 3 h).

**Superconductivity in CVD 1L-$NbSe_2$ films**

We next investigated the low-temperature electrical transport properties of air-stable, low-defect 1L-$NbSe_2$ films. For this purpose, we first developed a specialized set of fabrication methodologies tailored for air-sensitive 2D materials. As illustrated in Fig. 3a, following the synthesis by encapsulation epitaxy, we performed an oxidation-free transfer based on the electrical double layer force (EDL) transfer process[29], which enables the assembly of heterostructures with additional van der Waals materials and the larger-scale integration with other substrates and circuits (Supplementary Fig. 5). An example van der Waals heterostructure made of hBN, $NbSe_2$, graphene, and $MoS_2$ is shown in Supplementary Fig. 6, which simultaneously shows the Raman peaks of each functioning layer.

The four–probe measurements confirm that the 1L-$NbSe_2$ films are superconducting at low temperatures after transfer and device fabrication processes. Two types of device configurations were fabricated and characterized: 1) Top-contacted devices, where graphene/$NbSe_2$ heterostructures were transferred onto a new $SiO_2$/Si substrate by the EDL transfer method and contacted with top Al electrodes; and 2) Edge-contacted devices, using a similarly transferred graphene/$NbSe_2$ film, but instead contacted at the heterostructure edges via Al electrodes (See Supplementary Fig 5-7 and Method for fabrication details). As shown in Fig. 3d-e, for both methods, the measured resistance in the four-probe setup exhibits metallic behavior ($dR/dT > 0$) at high temperatures and a superconducting transition with critical temperature $T_c \approx 1$ K, which agrees with prior literature[14,16]. While both planar and edge-contacted devices eventually reach the superconducting state as the temperature is reduced, the edge-contacted device exhibits a sharper transition, with resistance starting to drop at $T_{onset} = 1.20$ K and dropping to zero at $T_{zero} = 0.85$ K. This difference is attributed to the contact configuration, as the Al electrodes contact both graphene and 1L-$NbSe_2$ in the edge-contacted case, allowing current to flow directly to the $NbSe_2$. For the planar-contacted device, by contrast, graphene serves as a normal layer between the Al and $NbSe_2$ films, introducing a series resistance and an inverse proximity effect that softens the superconducting transition[30]. We measure the differential resistance ($dV/dI$) and current-voltage ($I$-$V$) characteristics of the edge-contacted device by sweeping the DC bias current ($I_{DC}$) from -1 μA to 1 μA, as shown in Fig. 3f, and the critical current density ($J_c$) for 1L-$NbSe_2$ is inferred to be $3.7\times10^8$ A/m$^2$. Low hysteresis is observed, indicating minimal Joule heating and efficient thermal dissipation within the device[31].

Pseudo-four terminal measurements[32] further confirm that superconducting edge contacts with zero contact resistance are achieved in edge-contacted graphene/$NbSe_2$ devices at low temperature, as shown in Fig. 3g-i (see Methods for more discussions). Superconducting contacts are generally required when integrating 1L-SCs into quantum circuits to preserve high quality factors $Q_i$ (low loss). Encapsulation epitaxy enables the direct growth of 1L-graphene/$NbSe_2$ heterostructures with clean, bubble-free interfaces, where graphene serves as a medium for forming high-quality superconducting contacts, in addition to protecting $NbSe_2$ from ambient air.

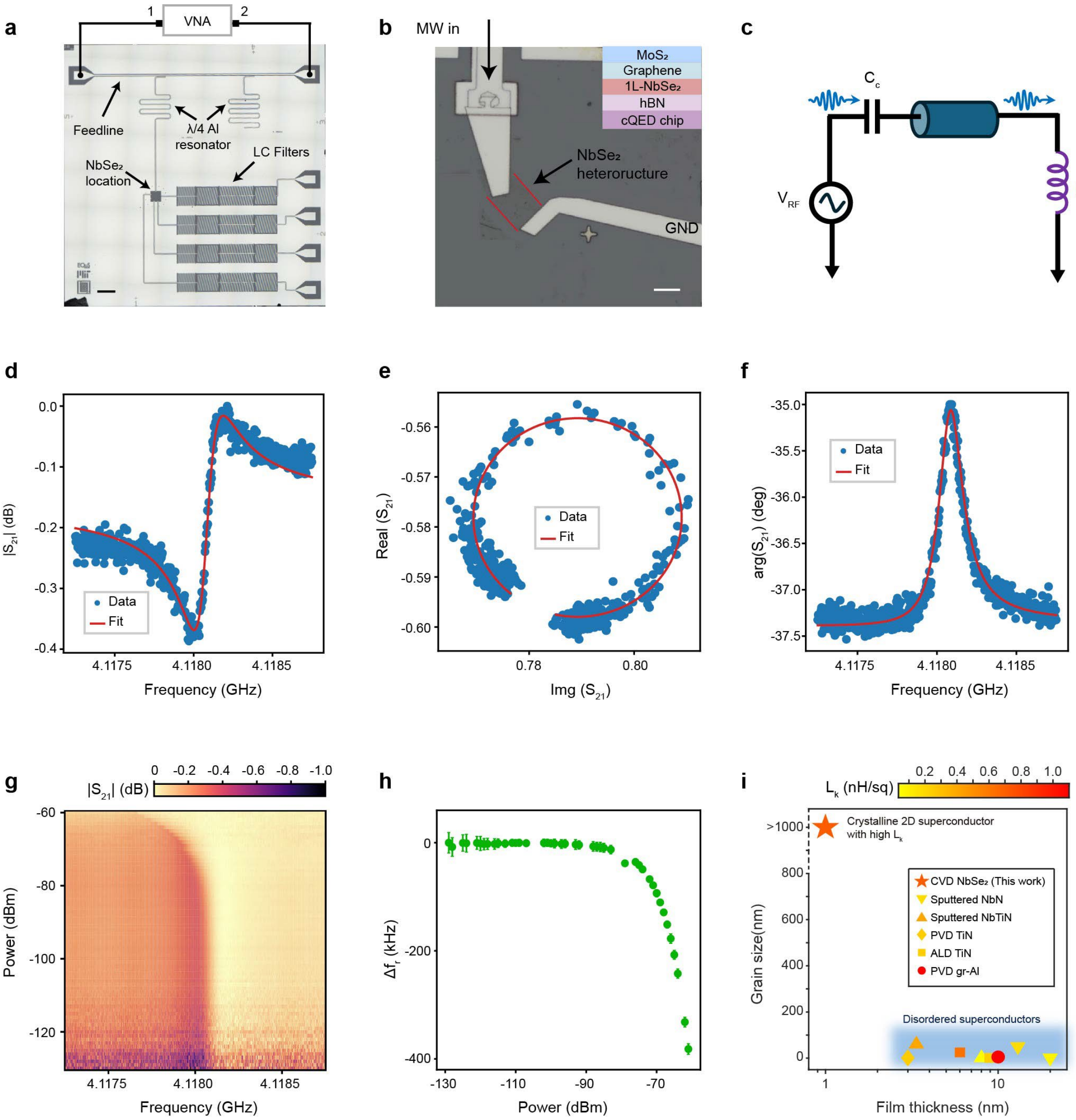

**Fig. 4 | Integrating 1L-$NbSe_2$ into a superconducting circuit. a,** Optical image of the superconducting circuit used to characterize the kinetic inductance of 1L-$NbSe_2$ films. Two λ/4 aluminum coplanar waveguide (CPW) resonators are capacitively coupled to a common feedline connected to a vector network analyzer. Scale bar, 0.2 mm. **b**, Optical image of 1L-$NbSe_2$ fully encapsulated by hexagonal boron nitride (bottom) and graphene and $MoS_2$ (top) integrated onto the Si chip. The $MoS_2$, graphene, and $NbSe_2$ are all monolayer continuous films synthesized via wafer-scale CVD techniques. The hBN is obtained by mechanical exfoliation. Scale bar, 10 μm. **c,** An equivalent circuit model for a λ/4 resonator terminated to ground via an inductor (1L-$NbSe_2$) and coupled capacitively to a feedline. **d,** Resonator spectroscopy (transmission coefficient $|S_{21}|^2$) of a λ/4 aluminum CPW terminated ground with 1L-$NbSe_2$ film, with an applied RF power of –110 dBm at the sample. **e,** Complex I–Q plot of the transmission coefficient ($S_{21}$), with blue dots showing the measured response, and the red trace the Lorentzian fit. A resonant frequency $f_r = 4.118$ GHz and internal quality factor $Q_i = 3.15\times10^4$ are extracted from

the circle fit (applied RF power of -110 dBm). **f,** Phase of $S_{21}$ as a function of frequency. **g,** Microwave power dependence of the $NbSe_2$-terminated λ/4 resonator. **h**, The resonant frequency shift, defined as $\Delta f_r = f_r(P_{in} = -130\ \mathrm{dBm}) - f_r(P)$, plotted as a function of the applied microwave power. The observed downward shift at higher powers indicates the onset of a nonlinear inductance in the superconducting film. **i,** Comparison of high kinetic inductance films synthesized by extensible production techniques, focusing on film thickness, crystallinity (grain size) and kinetic inductance ($L_k$).

### Integration with superconducting circuits

We demonstrate that encapsulation-epitaxially grown 1L-$NbSe_2$ can be integrated with conventional aluminum superconducting circuits. Fig. 4a and 4b show the superconducting circuit employed to characterize 2D superconductors in the microwave regime[33,34]. The measured $NbSe_2$ heterostructure comprises $MoS_2$/graphene/$NbSe_2$/hBN layers. All the components are continuous monolayer films grown by CVD (graphene, $NbSe_2$) or metal-organic chemical vapor deposition (MOCVD; $MoS_2$), except for hBN, which are exfoliated multilayer flakes (see Supplementary Fig. 5-6). The chip is cooled to 10 mK in a dilution refrigerator (see Supplementary Fig. 9 for the wiring schematic), and the microwave transmission coefficient $S_{21}$ is measured as a function of probe frequency using a vector network analyzer (VNA). Figure 4d-f shows the magnitude, phase and complex I-Q data of the 1L-$NbSe_2$-terminated λ/4-resonator. A Lorentzian fit[35] determines the resonant frequency of 1L-$NbSe_2$ terminated Al resonator $f_r$ = 4.118 GHz and the internal quality factor $Q_i = 3.15\times10^4$ at RF power of -110 dBm at the device. The $Q_i$ for single-photon limit is $2.29\times10^4$ (at -125 dBm; Supplementary Fig. 10).

To extract the kinetic inductance of 1L-$NbSe_2$ heterostructure, we compare the microwave response of the $NbSe_2$-terminated resonator with that of an all-aluminum λ/4-resonator with identical geometry. From an analytical model[34], we extract the sheet kinetic inductance of the CVD 1L-$NbSe_2$ to be 0.71 nH/□. Note that this value likely represents the lower bound for the kinetic inductance of CVD-grown 1L-$NbSe_2$ in isolation due to the presence of the graphene layer. In addition, we sweep the microwave power from -130 dBm to -60 dBm at the device (Fig. 4g) and observe a resonant frequency shift of 400 kHz, which can be attributed to the Kerr non-linearity of the film. Finally, we benchmark our CVD grown 1L-$NbSe_2$ film against other high kinetic inductance materials reported in the literature that are synthesized using wafer-scale-compatible approaches[36–43]. Current state-of-the-art inductors primarily rely on amorphous or polycrystalline films with nanometer-scale grain sizes (disordered superconductors). The grain size of 1L-$NbSe_2$ is > 1 μm, much better crystallinity than other materials such as NbN, NbTiN, TiN, granular aluminum (gr-Al). We note that the measured flake size in a partially grown 1L-$NbSe_2$ is ≈ 1 μm; however, since the growth is epitaxial, grain size can be much larger than 1 μm assuming the grain stitching is seamless. Therefore, the continuous, air-stable 1L-$NbSe_2$ film synthesized in this work is a highly crystalline monolayer superconductor with high kinetic inductance. This opens opportunities to construct quantum circuits based on wafer-scale superconducting 2D heterostructures using monolithic fabrication processes, making it well-suited for superconducting quantum circuits in applications such as metrology[44], sensing[45] and quantum computing[46].

In conclusion, the encapsulation epitaxy method developed in this work enables wafer-scale growth of monolayer superconducting films at the at the interface of a 2D vdW material (e.g., graphene, hBN) and a 3D substrate (e.g., $SiO_2$, $Si_3N_4$) with unprecedented air stability, demonstrating strong potential for use in fabricating superconducting quantum devices. The approach is applicable to other 2D material systems (Extended Data Fig. 9) and provides a new route to directly synthesize van der Waals heterostructures, particularly benefiting air-sensitive monolayers. We envision that this encapsulation epitaxy method will enable a broad class of monolayer materials for advancing technologies beyond quantum computing.

# References

1. Saito, Y., Nojima, T. & Iwasa, Y. Highly crystalline 2D superconductors. *Nat. Rev. Mater.* **2**, 16094 (2017).
2. Liu, X. & Hersam, M. C. 2D materials for quantum information science. *Nat. Rev. Mater.* **4**, 669–684 (2019).
3. Xi, X. *et al.* Strongly enhanced charge-density-wave order in monolayer $NbSe_2$. *Nat Nanotechnol* **10**, 765–769 (2015).
4. Peri, V., Ilani, S., Lee, P. A. & Refael, G. Probing quantum spin liquids with a quantum twisting microscope. *Phys Rev B* **109**, (2024).
5. Xi, X. *et al.* Ising pairing in superconducting $NbSe_2$ atomic layers. *Nat Phys* **12**, 139–143 (2016).
6. Wang, J. I. J. *et al.* Hexagonal boron nitride as a low-loss dielectric for superconducting quantum circuits and qubits. *Nat Mater* **21**, 398–403 (2022).
7. Antony, A. *et al.* Miniaturizing Transmon Qubits Using van der Waals Materials. *Nano Lett* **21**, 10122–10126 (2021).
8. Balgley, J. *et al.* Crystalline superconductor-semiconductor Josephson junctions for compact superconducting qubits. *Phys Rev Appl* **24**, 034016 (2025).
9. Cao, Y. *et al.* Quality Heterostructures from Two-Dimensional Crystals Unstable in Air by Their Assembly in Inert Atmosphere. *Nano Lett* **15**, 4914–4921 (2015).
10. Li, T. *et al.* Epitaxial growth of wafer-scale molybdenum disulfide semiconductor single crystals on sapphire. *Nat Nanotechnol* **16**, 1201–1207 (2021).
11. Zhu, J. *et al.* Low-thermal-budget synthesis of monolayer molybdenum disulfide for silicon back-end-of-line integration on a 200 mm platform. *Nat Nanotechnol* **18**, 456–463 (2023).
12. Xia, Y. *et al.* 12-inch growth of uniform $MoS_2$ monolayer for integrated circuit manufacture. *Nat Mater* **22**, 1324–1331 (2023).

13. Zhou, J. *et al.* A library of atomically thin metal chalcogenides. *Nature* **556**, 355–359 (2018).
14. Ugeda, M. M. *et al.* Characterization of collective ground states in single-layer $NbSe_2$. *Nat Phys* **12**, 92–97 (2016).
15. Naritsuka, M., Machida, T., Asano, S. *et al.* Superconductivity controlled by twist angle in monolayer $NbSe_2$ on graphene. *Nat. Phys.* **21**, 746–753 (2025).
16. Wang, H. *et al.* High-quality monolayer superconductor $NbSe_2$ grown by chemical vapour deposition. *Nat Commun* **8**, (2017).
17. Li, J. *et al.* Printable two-dimensional superconducting monolayers. *Nat Mater* **20**, 181–187 (2021).
18. Zhou, Z. J. et al. Stack growth of wafer-scale van der Waals superconductor heterostructures. *Nature* **621**, 499–505 (2023).
19. Fu, Z. *et al.* Van der Waals growth of monolayer transition metal dichalcogenide superconductors on ultra-flat graphene. *2d Mater* **12**, 015021 (2024).
20. Chen, C. W., Choe, J. & Morosan, E. electron systems. *Reports on Progress in Physics* **79**, (2016).
21. Cho, K. et al. Using controlled disorder to probe the interplay between charge order and superconductivity in $NbSe_2$. *Nat. Commun.* **9**, 2796 (2018).
22. Venables, J. A., Spiller, G. D. T. & Hanbucken, M. Nucleation and growth of thin films. *Reports on Progress in Physics* **47**, 399–459 (1984).
23. Venables, J. A. Atomic processes in crystal growth. *Surf Sci* **200–300**, 798–817 (1994).
24. Seebauer, E. G. & Allen, C. E. Estimating surface diffusion coefficients. *Prog Surf Sci* **49**, 265–330 (1995).
25. Wickramaratne, D., Khmelevskyi, S., Agterberg, D. F. & Mazin, I. Ising superconductivity and magnetism in $NbSe_2$. *Phys. Rev. X* **10**, 041003 (2020)..
26. Moon, D. *et al.* Hypotaxy of wafer-scale single-crystal transition metal dichalcogenides. *Nature* **638**, 957–964 (2025).
27. Lee, J., Ahn, G., Shim, J. *et al.* Optical separation of mechanical strain from charge doping in graphene. *Nat Commun* **3**, 1024 (2012).
28. Wang, Y. et al. Modulation doping via a two-dimensional atomic crystalline acceptor. *Nano Lett.* **20**, 8446–8452 (2020).
29. Zheng, X. *et al.* Electrostatic-repulsion-based transfer of van der Waals materials. *Nature* **645**, 906–914 (2025).
30. Bagwell, P. F. Suppression of the Josephson current through a narrow, mesoscopic, semiconductor channel by a single impurity. *Phys. Rev. B* **46**, 12573–12586 (1992).
31. Tinkham, M., Free, U., Lau, N. & Markovic, N. Hysteretic I-V curves of superconducting nanowires. *Phys Rev B Condens Matter Mater Phys* **68**, (2003).
32. Sinko, M. R. *et al.* Superconducting contact and quantum interference between two-dimensional van der Waals and three-dimensional conventional superconductors. *Phys Rev Mater* **5**, (2021).

33. Tanaka, M. *et al.* Superfluid stiffness of magic-angle twisted bilayer graphene. *Nature* **638**, 99–105 (2025).
34. Zaman, S. et al. Kinetic Inductance of Few-Layer $NbSe_2$ in the Two-Dimensional Limit. http://arxiv.org/abs/2511.08466 (2025).
35. Probst, S., Song, F. B., Bushev, P. A., Ustinov, A. V. & Weides, M. Efficient and robust analysis of complex scattering data under noise in microwave resonators. *Review of Scientific Instruments* **86**, (2015).
36. Frasca, S. *et al.* NbN films with high kinetic inductance for high-quality compact superconducting resonators. *Phys Rev Appl* **20**, (2023).
37. Bretz-Sullivan, T. M. *et al.* High kinetic inductance NbTiN superconducting transmission line resonators in the very thin film limit. *Appl Phys Lett* **121**, (2022).
38. Amin, K. R. *et al.* Loss mechanisms in TiN high impedance superconducting microwave circuits. *Appl Phys Lett* **120**, (2022).
39. Winkel, P. *et al.* Implementation of a Transmon Qubit Using Superconducting Granular Aluminum. *Phys Rev X* **10**, (2020).
40. Niepce, D., Burnett, J. & Bylander, J. High Kinetic Inductance NbN Nanowire Superinductors. *Phys Rev Appl* **11**, (2019).
41. Shearrow, A. *et al.* Atomic layer deposition of titanium nitride for quantum circuits. *Appl Phys Lett* **113**, (2018).
42. Samkharadze, N. *et al.* High-Kinetic-Inductance Superconducting Nanowire Resonators for Circuit QED in a Magnetic Field. *Phys Rev Appl* **5**, (2016).
43. Coumou, P. C. J. J. *et al.* Microwave properties of superconducting atomic-layer deposited TiN films. *IEEE Transactions on Applied Superconductivity* **23**, (2013).
44. Shaikhaidarov, R. S. *et al.* Quantized current steps due to the a.c. coherent quantum phase-slip effect. *Nature* **608**, 45–49 (2022).
45. Szypryt, P., Bennett, D.A., Fogarty Florang, I. *et al.* Kinetic inductance current sensor for visible to near-infrared wavelength transition-edge sensor readout. *Commun Eng* **3**, 160 (2024)
46. Rieger, D. *et al.* Granular aluminium nanojunction fluxonium qubit. *Nat Mater* **22**, 194–199 (2023).

# Methods

**Encapsulation epitaxy for growing monolayer $NbSe_2$**

The encapsulation epitaxy is conducted using a hot-wall CVD system at ambient pressure with a fused quartz tube. We used niobium oxide film (obtained by first depositing 10 nm Nb on $SiO_2$/Si substrate by magnetron sputtering, and then oxide in air at 500℃ for 5 hours) and Selenium powder as growth precursor. Prior to growth, the 2D/3D hybrid substrate (e.g. graphene/$SiO_2$) is prepared by transferring or exfoliating a 2D layer onto $SiO_2$/Si (285 nm $SiO_2$) substrate. The 2D layers used in this work are either a layer of CVD grown monolayer graphene film transferred from copper substrate or directly exfoliated h-BN flakes. Supplementary Fig. 2 shows the set-up of the CVD reaction chamber. The 2D/3D hybrid substrate and niobium oxide film are placed face-to-face at the center of the hot wall furnace, with a distance of 1-3 mm between each other. The face-to-face design ensures uniform precursor delivery to the growth substrate[47], enabling uniform film growth. The NaCl (1.8 g) is placed at the upstream in the middle between furnace center and the Se boat. The Se powder (2.5 g) (99.5%, Sigma-Aldrich) is placed at upstream edge. 60 sccm (cubic centimeters per minute) Ar and 6 sccm $H_2$ are used as carrier gas. The furnace is heated up to 950 °C in 15 min and maintained at that temperature for 9 minutes for the growth of $NbSe_2$ at 2D-$SiO_2$ interface. Then the top cover of the furnace is opened to allow fast cooling of the sample.

It should be emphasized that the implementation of "face-to-face" precursor delivery is made possible by a key innovation in NaCl vapor delivery. At the growth temperature of 950 °C, niobium oxide alone cannot evaporate; the presence of NaCl is essential, as it lowers the melting point of niobium oxide and facilitates its evaporation. However, when NaCl solids are physically mixed with the metal oxide precursor, NaCl tends to evaporate significantly faster than the metal precursor and becomes depleted before the growth is completed, which results in an unstable vapor environment and hinders the ability to achieve continuous monolayer growth. To overcome this problem, we developed a vapor-phase NaCl based salt delivery method. As illustrated in Supplementary Fig. 2a, NaCl is placed upstream in a lower-temperature zone, physically separated from niobium oxide. Upon heating (≈ 750 – 850 °C), NaCl melts and evaporates at a nearly constant rate, which is determined by the local temperature, throughout the growth. When NaCl vapor is carried by the Ar/$H_2$ carrier gas to the growth region, it interacts with niobium oxide that is placed face-to-face to the substrate, effectively lowering its melting point and enabling continuous and controlled evaporation of the niobium oxide precursor. This approach ensures a steady Nb precursor flux, which is critical for the uniform and reproducible growth of continuous 1L-$NbSe_2$ films.

**Selective-area growth**

The selective growth of $NbSe_2$ demonstrated in Extended Data Fig. 8 is achieved using patterned graphene on a $SiO_2$ substrate. A key requirement for such selective-area growth is the suppression of unwanted nucleation on the exposed $SiO_2$ regions. We found that this suppression is strongly dependent on the local mass flux of metal precursors. Under high precursor concentrations, thick $NbSe_2$ flakes tend to form even on $SiO_2$, likely due to elevated supersaturation that facilitates nucleation. However, when the Nb precursor concentration is reduced, such parasitic growth on $SiO_2$ is effectively suppressed. Importantly, growth at the 2D/3D interface still persists, as the graphene edges act as favorable nucleation sites. In practice, this reduction in precursor concentration is achieved by increasing the distance between the niobium oxide source and the substrate from 1 mm to 2–3 mm, while keeping other growth parameters unchanged. The graphene patterns are defined using standard lithography and etching processes, allowing full customization of the pattern and shape.

**DFT calculations**

The first-principles calculations based on density functional theory (DFT) are performed using the Vienna ab initio simulation package (VASP)[48]. The exchange-correlation interactions are incorporated by the generalized gradient approximation (GGA) in the form of Perdew-Burke-Ernzerhof (PBE). Core and valence electrons are treated by projector augmented wave (PAW) method and plane-wave basis functions, respectively. The nudged elastic band[49] method is used to determine the diffusion barriers.

**Calculation of diffusion coefficient** $D_s$ **and surface residence time** $\tau$

The diffusion coefficient follows an Arrhenius form $D_s = D_0 \cdot e^{-E_{diff}/k_B T}$, where $D_0$ is the diffusion pre-factor, $E_{diff}$ is the surface diffusion barrier extracted from DFT calculations, $k_B$ is the Boltzmann constant, and $T$ is the growth temperature[27]. The prefactor $D_0$ was taken to be $1\times10^{-3}$ cm$^2$/s, a typical value for surface diffusion of adatoms on solid surfaces[28]. The surface residence time $\tau$ was estimated using: $\tau = \nu^{-1} \cdot e^{E_{ads}/k_B T}$, where $E_{ads}$ is the adsorption energy (approximated as the magnitude of DFT-calculated binding energy), and $\nu$ is a characteristic surface vibration frequency , taken to be ≈ 10 THz that is commonly used for thermally activated surface processes[26,27,54]. The growth temperature is $T$ = 950 °C. For $NbSe_2$ on $SiO_2$, with $E_{diff} \approx 1.0$ eV and $E_{ads} \approx 2.6$ eV, we obtained $D_s \approx 7.6 \times 10^{-8}$ cm$^2$/s, $\tau \approx$ 5.2 ms. For $NbSe_2$ on graphene, where $E_{diff} \approx 0.3$ eV and $E_{ads} \approx 1.9$ eV, we obtained $D_s \approx 5.8 \times 10^{-5}$ cm$^2$/s, $\tau \approx 6.8 \times 10^{-3}$ ms. These contrasting values illustrate the interplay between residence time and surface mobility, and help rationalize the substrate-dependent growth behaviors observed in our experiments.

**Substrate dependent growth test for understanding the growth mechanism**

In addition to $SiO_2$, we further evaluated the applicability of encapsulation epitaxy using 2D template on other amorphous 3D substrates, including $Si_3N_4$, $HfO_2$ and $Al_2O_3$. This assessment combined both DFT calculations and experimental growth tests, as presented in Extended Data Figs. 3 and 4. It turned out that encapsulation epitaxy can be successfully performed using $Si_3N_4$. However, it was not suitable for $HfO_2$ and $Al_2O_3$, on which only thick flakes and black dense film were observed. DFT calculations (Extended Data Fig. 3) indicate that $Si_3N_4$ exhibits $E_{adsorb}$ and $E_{diff}$ values comparable to those $SiO_2$, supporting both nucleation and lateral growth condition required by encapsulation epitaxy. In contrast, $HfO_2$ and $Al_2O_3$ show significantly higher magnitude of $E_{diff}$, which hinder the ability of 2D template to mediate the surface diffusion barrier. These findings are consistent with both our theoretical understanding and experimental observations.

**Shape of the partially grown $NbSe_2$ flakes**

It should be noted that the apparent shape of the partially grown $NbSe_2$ shown in Fig. 1b does not reflect the true morphology of individual $NbSe_2$ crystals. As revealed in Supplementary Fig. 1, each optical-domain $NbSe_2$ region is actually composed of many small crystalline flakes with grain size of ≈ 1 μm (since the growth is epitaxial, the real grain size can be much larger than 1 μm assuming the grain stitching is seamless). The larger shapes observed in the optical image are defined by graphene wrinkles. This suggests that $NbSe_2$ growth can be initiated at the wrinkle sites, which is reasonable considering the fact that the wrinkle location has larger gap size and thus serve as the entrance for the precursors to enter the van der Waals gap. Another entry point for precursors is the edge of the graphene layers.

**Verification of growth at 2D/3D interface – Cryogenic dry pick-up test**

The unusual growth behavior at 2D/3D interface in encapsulation epitaxy is evidenced by the cryogenic pickup test and Raman characterization shown in Extended Data Fig. 2. In this test, a cryogenic dry pick-up technique based on cold PDMS (<−90 °C) was utilized to pick up the as-grown hBN/$NbSe_2$ from growth substrate ($SiO_2$). The pickup process is detailed in a previous work[51]. It turned out that only the hBN was picked up while $NbSe_2$ stays on the original substrate, which means hBN is on top of $NbSe_2$. This, combined with the STEM cross section image and superior air-stability, confirms the 1L-$NbSe_2$ are grown at the 2D/3D interface, instead of on top of the 2D surface.

**Oxidation-free, wafer-scale transfer of $NbSe_2$ by electrical double layer force**

As shown in the cryogenic pick-up test, we found it challenging for dry transfer methods to detach the CVD grown 1L-$NbSe_2$, a difficulty that also applies to many other CVD-grown 2D materials. To enable device fabrication and circuit integration, we developed oxidation-free, wafer-scale wet transfer method based on Electrical double layer (EDL) forces utilizing ammonia solutions. A detailed discussion of the EDL transfer method is provided in our recent work[29]. The key modification in the present study is that both the detachment from the growth substrate and attachment to the target substrate are conducted in an inert environment, as the detachment temporarily exposes the bottom surface of $NbSe_2$ to ambient, making it vulnerable to oxidation.

The transfer process began by spin-coating the graphene/ $NbSe_2$ stack (also applicable to hBN/ $NbSe_2$) with PMMA A4 (molecular weight 950,000) at 3000 rpm for 60 seconds. The sample was then baked at 70 °C for 5 minutes. Both spin-coat and bake steps are performed in air, as the top graphene layer provides sufficient protection for the $NbSe_2$. The PMMA/graphene/$NbSe_2$ film is subsequently transferred into a nitrogen box (oxygen-free but not water-free, due to the presence of ammonia) for delamination from the growth substrate and transfer to the target substrate. Inside the nitrogen box, the PMMA/graphene/$NbSe_2$/$SiO_2$/Si was slowly put into the ammonia solution surface that enabled immediate detachment of $NbSe_2$ from $SiO_2$ by EDL force. The detached PMMA/graphene/$NbSe_2$ was rinsed with water, transferred from the water surface onto the target substrate, and subsequently dried naturally in air for approximately one hour. The sample was then taken out of glove box for PMMA removal. The as-transferred $NbSe_2$ show minimal damage, as confirmed by optical microscopy, Raman spectroscopy and preservation of superconductivity.

**Further discussions about superconducting edge contact**

Figure 3g shows the measurement configuration where we measure the resistance of the electrodes and the device without incorporating line resistance. The differential resistance (d$V$/d$I$) of the two terminals goes to zero (within instrument noise) below the critical current transition, which indicates the contact resistance (or any residual resistance at the interface of Al-$NbSe_2$) is very low when the film is superconducting. The line cut from Fig. 3h at zero bias current shows the resistance becomes zero (Fig. 3i) when 1L-$NbSe_2$ is superconducting, which confirms superconducting transparent contact between the Al electrodes and 1L-$NbSe_2$.

**Material characterization**

Raman spectra were measured using WITec alpha300 apyron Confocal Raman with an excitation wavelength of 532 nm and a grating of either 300 or 1800 grooves/mm. AFM images were taken using Asylum Cypher VRS in tapping mode. SEM images were obtained using a high-resolution SEM (ZEISS Merlin; 3 kV) with an in-Lens detector.

XPS spectra were acquired using a PHI VersaProbe II system equipped with a monochromatic Al Kα source (1,486.6 eV) and a 200 μm spot size. The measurements were performed at a gun power of 50 W and an operation voltage of 15 kV. To neutralize surface charging during data acquisition, the sample was flooded using electron and argon ion guns. All spectra were calibrated against the C 1s peak at 284.8 eV.

TEM samples used for cross-sectional STEM images were prepared by Helios 460F1 Dual Beam milling. The lamella was thinned to 50 - 100 nm thickness at an accelerating voltage of 30 kV with a current decreasing from 0.78 nA to 40 pA, followed by a fine polish at lower voltage with a current of 21 pA to clean the sample surface.

The cross-sectional HAADF-STEM images were captured with a Thermo Fisher Scientific Titan Themis 80-300 aberration corrected S/TEM at 300 kV. The HAADF-STEM images were post processed using DigitalMicrograph® software (Gatan, Inc.). To filter the image noise, the lattice contrast of HAADF-STEM images were enhanced by Gaussian smooth and band-pass filter. EDX mappings were acquired with a Super-X detector (4 silicon drift detectors) at 300 kV. Nb and Se signals are selected from L line to increase the intensity.

The atomic lattice imaging in Fig. 2a was conducted using a high-resolution STEM (Titan Themis Z G3 Cs-Corrected S/TEM; 60 kV). The SAED images were obtained using multipurpose transmission electron microscope (FEI Tecnai G2 Spirit TWIN; 120 kV). The graphene/$NbSe_2$ freestanding samples were prepared by transferring as-grown graphene/$NbSe_2$ from $SiO_2$/Si substrate onto a TEM grid, which is enabled by electrostatic repulsion force in ammonia solution (EDL transfer)[29].

**CDW characterization**

The optical detection of CDW in $NbSe_2$ was conducted using temperature dependent Raman spectroscopy measurements, which are performed on a Horiba Labram HR Evolution spectrometer equipped with a liquid nitrogen cooled CCD and 600 l/mm and 1800 l/mm spectrometer gratings for medium and high-resolution spectra, respectively. All measurements were performed using an excitation wavelength of $\lambda = 532$ nm with an incident power of approximately 100 μW on the sample. Bragg grate notch filters were used to filter the elastic line permitting measurements down to 5 $cm^{-1}$ energy transfer. The incident excitation laser is linearly polarized and passed through an achromatic half wave plate (Thorlabs) placed before a 50x objective (Olympus) to vary the angle of incident polarization on the sample, with a $1/e^2$ focus spot diameter of approximately 4 microns. The scattered light polarization is resolved into linear parallel (X') and perpendicular (Y') components using a linear polarizer placed before the entrance slit of the spectrometer. Variable temperature measurements are performed using a Montana Instruments S50 closed-cycle optical cryostat.

The charge density wave (CDW) transition temperature, $T_{CDW}$, is determined from the intensity of the amplitude mode $I_A$, which is reported in the literature[3]. Raman spectra $S_T(\omega)$ collected at temperature $T$ were first normalized by a reference spectrum $S_0(\omega)$ measured at a temperature $T_0 > T_{CDW}$ . Subsequently, 1 was subtracted from this ratio to eliminate temperature-independent contributions unrelated to the CDW transition. The integrated intensity of the amplitude mode was then extracted as a function of temperature (Fig. 2e). The transition temperature $T_{CDW}$ was identified as the point where $I_A$ crosses zero. The solid curves in Fig. 2e and Extended Data Fig. 8c are fits based on mean-field theory, used as guides to the eye (see Supplementary Information part of reference[3] for the equation being used).

**Device fabrication and characterization**

After the 1L-$NbSe_2$ based heterostructure being transferred to the target substrate, it goes through standard device fabrication process in ambient conditions. It is first patterned and etched into a designed shape using Electron Beam Lithography (EBL) and Reactive Ion Etching (RIE). After removing the resist, another EBL step is conducted to define metal contact patterns. After developing the pattern, 1L-$NbSe_2$ based heterostructure goes through another etching to expose the edges, followed by 160 nm Al deposition. In-situ ion milling is not required in this process as the graphene serves as an ideal layer for constructing edge contact, which overcomes the challenge of oxidation of exposed $NbSe_2$ edges. After completing the metal deposition and lift-off processes, the sample is mounted and cooled down to a base temperature of 10 mK using a dilution refrigerator (Bluefors LD400).

During most of the device fabrication process, the $NbSe_2$ can be handled, transported and stored in ambient conditions without post-growth encapsulation layer, a unique advantage provided by encapsulation epitaxy. For devices that involve the step of metal deposition, an additional post-growth protection layer was applied to ensure $NbSe_2$ is not damaged by the high energy metal atoms (we use monolayer $MoS_2$ as it is readily accessible to us). Supplementary Fig. 7 shows a comprehensive comparison of graphene/$NbSe_2$ with and without extra capping layer during the fabrication steps.

For the pseudo-four-terminal measurement in Fig. 3g-i, the measured resistance represents the total series resistance of 1L-$NbSe_2$, the $NbSe_2$-Aluminum interface and aluminum leads between the interface and voltage probe. It drops to zero at low temperature, confirming the superconducting edge contact is achieved with zero resistance between 1L-$NbSe_2$ and conventional 3D superconductor (Al).

**Transport measurement setup**

Each DC line inside the dilution refrigerator is equipped with cryogenic filters at 4K stage (RC $\pi$-filter with a cutoff frequency of 53kHz) and a low-pass filter (RC-RF filter from QDevil) at the mixing chamber plate to get rid of high frequency noise. Supplementary Fig. 8 shows the wiring diagram for DC transport measurement from room temperature to the device. For measuring the resistance of our devices as a function temperature, an AC signal is provided by an SRS 860 lock-in amplifier and is applied through the BNC cables. An SRS 560 low-noise voltage amplifier (×1000) is used to amplify the signal, which is then read by the lock-in amplifier. For differential measurements, both AC and DC signals are applied with the current bias scheme using the lock-in for small AC current and a QDAC, which is a multi-channel, ultra-low-noise voltage source/digital-to-analog converter (from Qdevil, Denmark), as a DC voltage source. Two resistors with resistance much larger compared to the device's resistance are used to create current bias from voltage sources: $R_{ac}$ = 10 MΩ and $R_{dc}$ = 100 kΩ. The amplified voltages are measured by the lock-in input channel and Keithley 2100 digital multimeter (DMM). All instruments mentioned here, and the fridge share a common ground.

**Microwave measurement setup**

The microwave measurement setup is shown in Supplementary Fig. 9 where the device is mounted at the mixing chamber stage with via coax cables. The input-output lines are attenuated by 20 dB at the 4K stage, 10 dB at the still stage (1K), and 40 dB at the mixing chamber stage to minimize thermal noise from higher temperature stages. The output line is filtered with 3GHz high-pass and 12 GHz low-pass filters at the mixing chamber stage. We have a Josephson traveling wave parametric amplifier (TWPA) at the mixing chamber stage, and low noise high-electron-mobility transistor (HEMT) amplifiers at the 4K stage and room-temperature stages to amplify the output signal. A vector network analyzer (VNA) from Keysight (N5232A) is used to send and receive microwave signals.

# Additional References

47. Xue, G. et al. Large-area epitaxial growth of transition metal dichalcogenides. *Chem. Rev.* **124**, 9785–9865 (2024).
48. Kresse, G. & Furthmüller, J. Efficiency of ab-initio total energy calculations for metals and semiconductors using a plane-wave basis set. *Comput. Mater. Sci.* **6**, 15–50 (1996).
49. Sheppard, D., Terrell, R. & Henkelman, G. Optimization methods for finding minimum energy paths. *J. Chem. Phys.* **128**, 134106 (2008).
50. Redhead, P. A. Thermal Desorption of Gases. *Vacuum* **12**, 203–211 (1962).
51. Frank Zhao, S. Y. *et al.* Time-reversal symmetry breaking superconductivity between twisted cuprate superconductors. *Science (1979)* **382**, 1422–1427 (2023).
52. Zhang, Y. et al. Edge-epitaxial growth of 2D $NbS_2$-$WS_2$ lateral metal-semiconductor heterostructures. *Adv. Mater.* **30**, 1803665 (2018).
53. Su, J. et al. Sub-millimeter-scale monolayer p-type H-phase $VS_2$. *Adv. Funct. Mater*. **30**, 2000240 (2020).
54. You, J. *et al.* Salt-Assisted Selective Growth of H-phase Monolayer $VSe_2$ with Apparent Hole Transport Behavior. *Nano Lett* **22**, 10167–10175 (2022).

# Data availability

All data needed to evaluate the conclusions herein are present in the Article.

# Acknowledgements

We thank Dr. Batyr Ilyas for the valuable discussions. This work was carried out in part through the use of MIT.nano's facilities. X.Z., S.Z., S.P., J.I.W., W.D.O. and J.K. acknowledge the support by the US Army Research Office grant number W911NF2210023. S.Z., S.P., J.I.W., and W.D.O. acknowledge the support from the National Science Foundation for grant number 2412810. S.Z. acknowledges support from the Faculty for the Future Fellowship from the Schlumberger Foundation. K.Z., T.H.Y. and J. K. acknowledge the support by the National Science Foundation under Award No. 2527588. C.A.O., L.G.P.M. and R.C. acknowledge the support by the U.S. Department of Energy, Office of Science National Quantum Information Science Research Center's Co-design Center for Quantum Advantage (C2QA) under contract number DE-SC0012704. C2QA participated) in this research. Y.Z.

and J.K. acknowledge the support by the Air Force Office of Scientific Research under award number FA2386-24-1-4049. Z.W. and J.K. acknowledge the support by the Semiconductor Research Corporation Center 7 in JUMP 2.0 (award no. 145105-21913). T.Z. and J.K. acknowledge the support by the US Department of Energy (DOE), Office of Science, Basic Energy Sciences (BES) under award DE-SC0020042. K.Y.M., Z. H. and J.K. acknowledge the support from the US Army DEVCOM ARL Army Research Office through the MIT Institute for Soldier Nanotechnologies under Cooperative Agreement number W911NF-23-2-0121. The cross-sectional TEM studies were performed using the facilities in the UConn/Thermo Fisher Scientific Center for Advanced Microscopy and Materials Analysis (CAMMA). S.L. and P.K. acknowledge support from DMR-2105048 (NSF). S.P. acknowledge the support by the National Research Foundation of Korea (Grant No. RS-2025-02317602). S.V., K. T., S. M. acknowledge the support by the Department of the Air Force under Air Force Contract No. FA8702-15-D-0001 and FA8702-25-D-B002.
The views and conclusions contained herein are those of the authors and should not be interpreted as necessarily representing the official policies or endorsements of the US Government, the National Science Foundation, or the Department of the Air Force.

# Contributions

J.K., J.I.W., and W.D.O. supervised the project. X.Z. and S.Z. conceived the experiments. X.Z. developed the encapsulation epitaxy growth method. S.Z. performed superconducting device fabrication and measurements. X.Z., K.Z., and Z.W. carried out material synthesis and characterization. C.O. and L.G.P.M. conducted low temperature Raman measurements, under supervision of R.C. H.X. performed DFT calculations. F.L. conducted cross-section STEM imaging. Z.W. and Y.Z. provided CVD graphene films. S.L. conducted cryogenic dry pick-up tests. T.Z. contributed the graphic illustration. T.Y. and S.P. contributed to device fabrication. J.W. and Z.H. contributed to data analysis. S.V., K.T., and S.M. prepared Nb films. X.Z., S.Z., J.I.W. and J.K. wrote the manuscript; all authors read and revised the manuscript.

# Competing interests

The authors declare no competing interests.

# Extended Data Figures

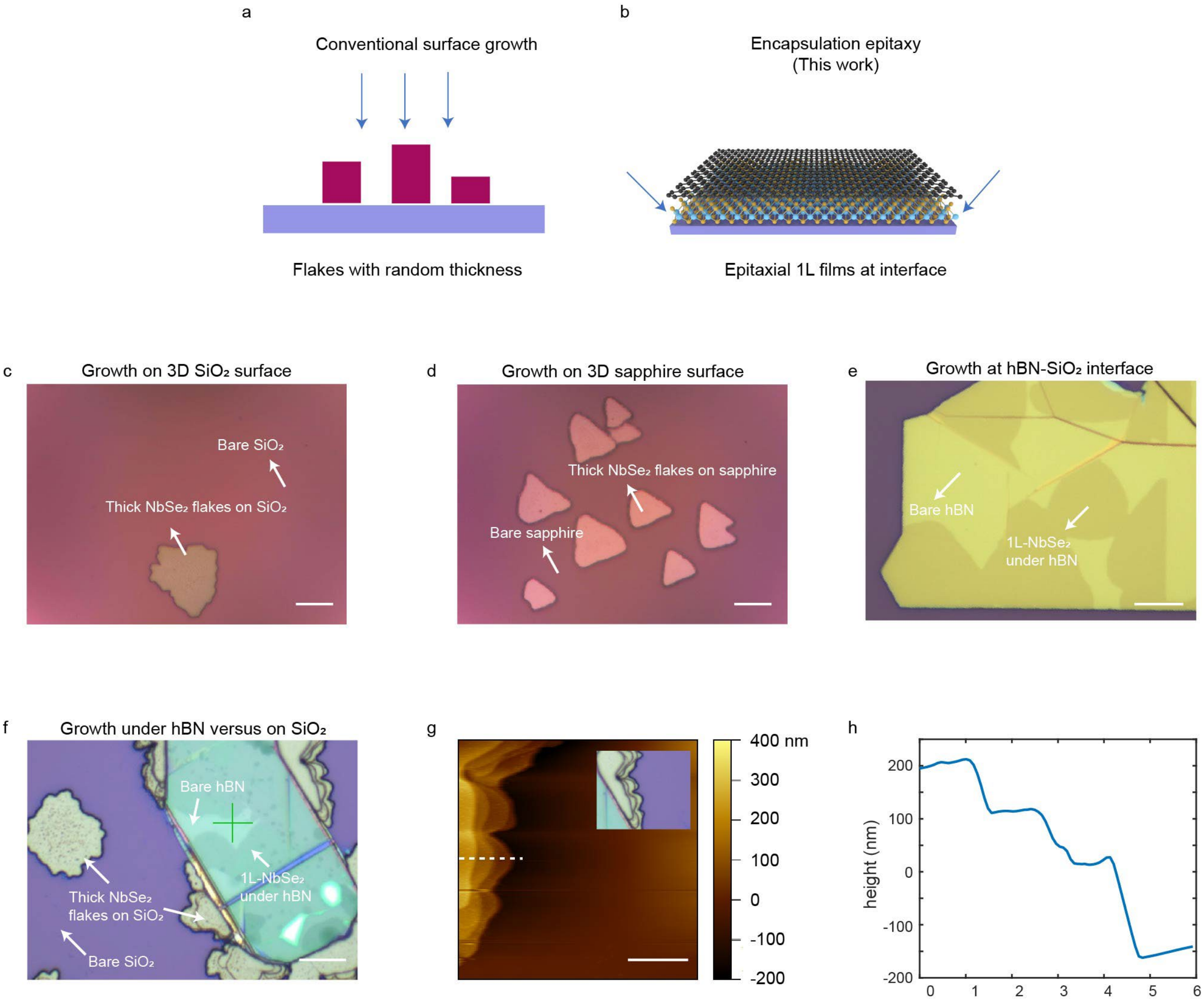

**Extended Data Fig. 1 | Comparison between conventional surface growth versus encapsulation epitaxy. a-b,** Schematic showing the difference between (a) conventional surface growth on 3D surface and (b) Encapsulation epitaxy at 2D-3D interface. **c-e,** Optical image showing the growth of $NbSe_2$ on (c) $SiO_2$ substrate surface, (d) Sapphire substrate surface, and (e) exfoliated hBN flakes on $SiO_2$. Scale bars, 10 µm. **f**, Additional example showing the difference between $SiO_2$ surface and hBN-$SiO_2$ interface. This growth utilizes a high Nb precursor concentration, enabling the simultaneous observation of island growth on $SiO_2$ surface (conventional surface growth) and 1L-$NbSe_2$ growth at hBN-$SiO_2$ interface (encapsulation epitaxy). Scale bar, 10 µm. **g,** AFM image measuring the thickness of the thick $NbSe_2$ flake on $SiO_2$ surface. Inset, optical image showing the measured location). Scale bar, 1 µm. **h,** Corresponding height profile of the AFM image shown in panel **g** (dashed white line).

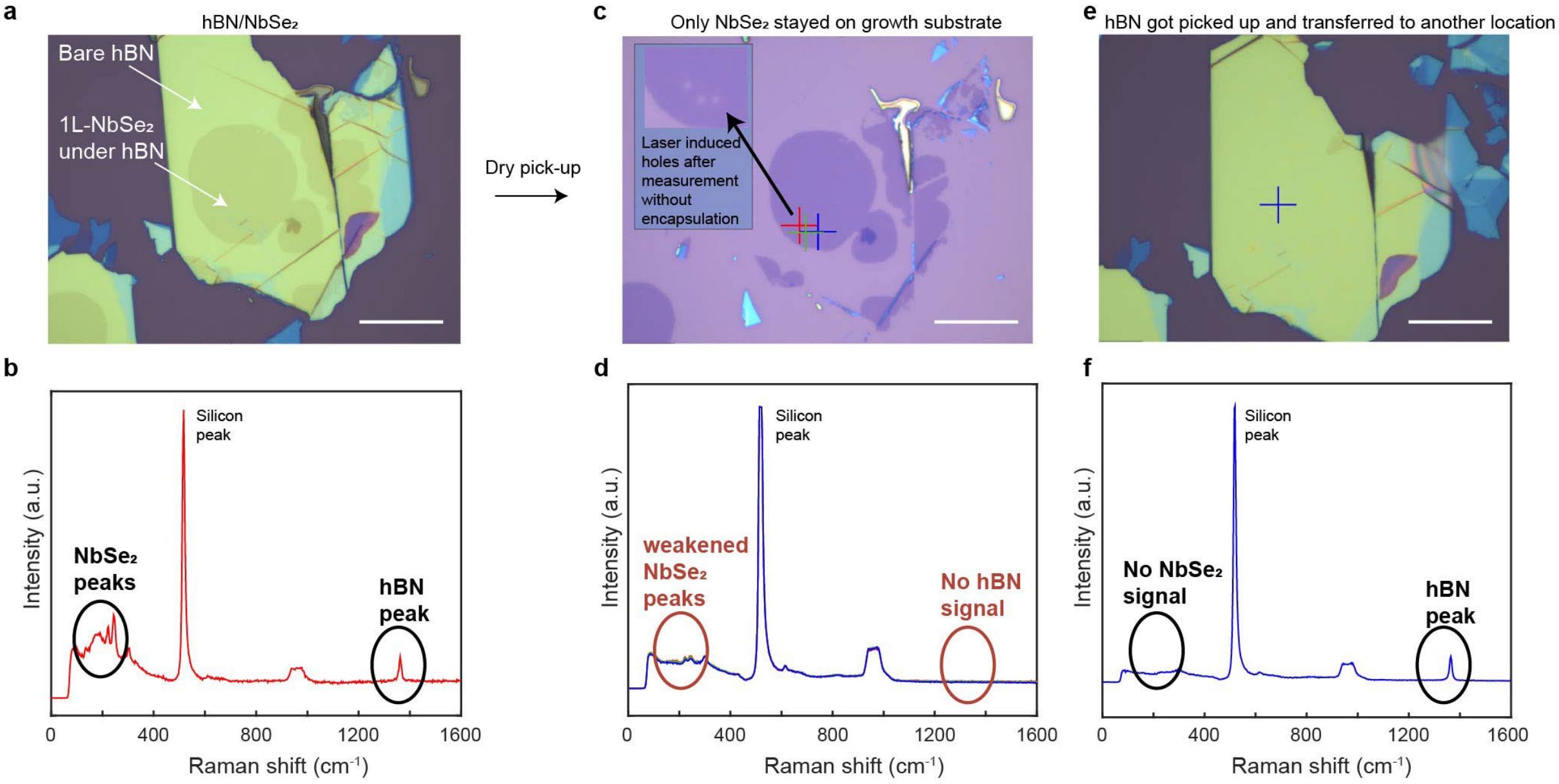

**Extended Data Fig. 2 | Cryogenic dry pick-up test demonstrating that $NbSe_2$ growth occurred exclusively at 2D/3D interface, with no growth observed on the exposed 2D surface. a,** Optical image showing the 1L-$NbSe_2$ grown by encapsulation epitaxy at the hBN/$SiO_2$ interface. **b,** Raman spectra of as-grown hBN/$NbSe_2$ from panel **a**. **c,** Optical image of the same sample shown in panel **a** after cryogenic dry pick-up (see Methods). The $NbSe_2$ remains on the growth substrate while the hBN flakes were lifted off, confirming that $NbSe_2$ was grown beneath hBN at the 2D/3D interface. After removal of the hBN layer, further laser exposure will cause damage to the unprotected $NbSe_2$ surface. The inset shows the optical image of laser-induced holes on exposed $NbSe_2$ surface, underscoring the importance of 2D encapsulation. **d,** Raman spectra of the exposed $NbSe_2$ after removal of the top hBN layers. The $NbSe_2$ peaks are markedly weakened due to the absence of hBN encapsulation during laser exposure. The three spectra correspond to measurements taken at the positions indicated by the three crosses in panel **c**. **e,** Optical image of the hBN flake being picked up in a and transferred onto another new substrate. **f,** Raman spectra measuring the cross-marked location in panel **e**. No $NbSe_2$ signal was detected, together with the uniform color contrast under the optical microscope image, indicating no growth happened on top of the hBN surface. Scale bars, 20 µm.

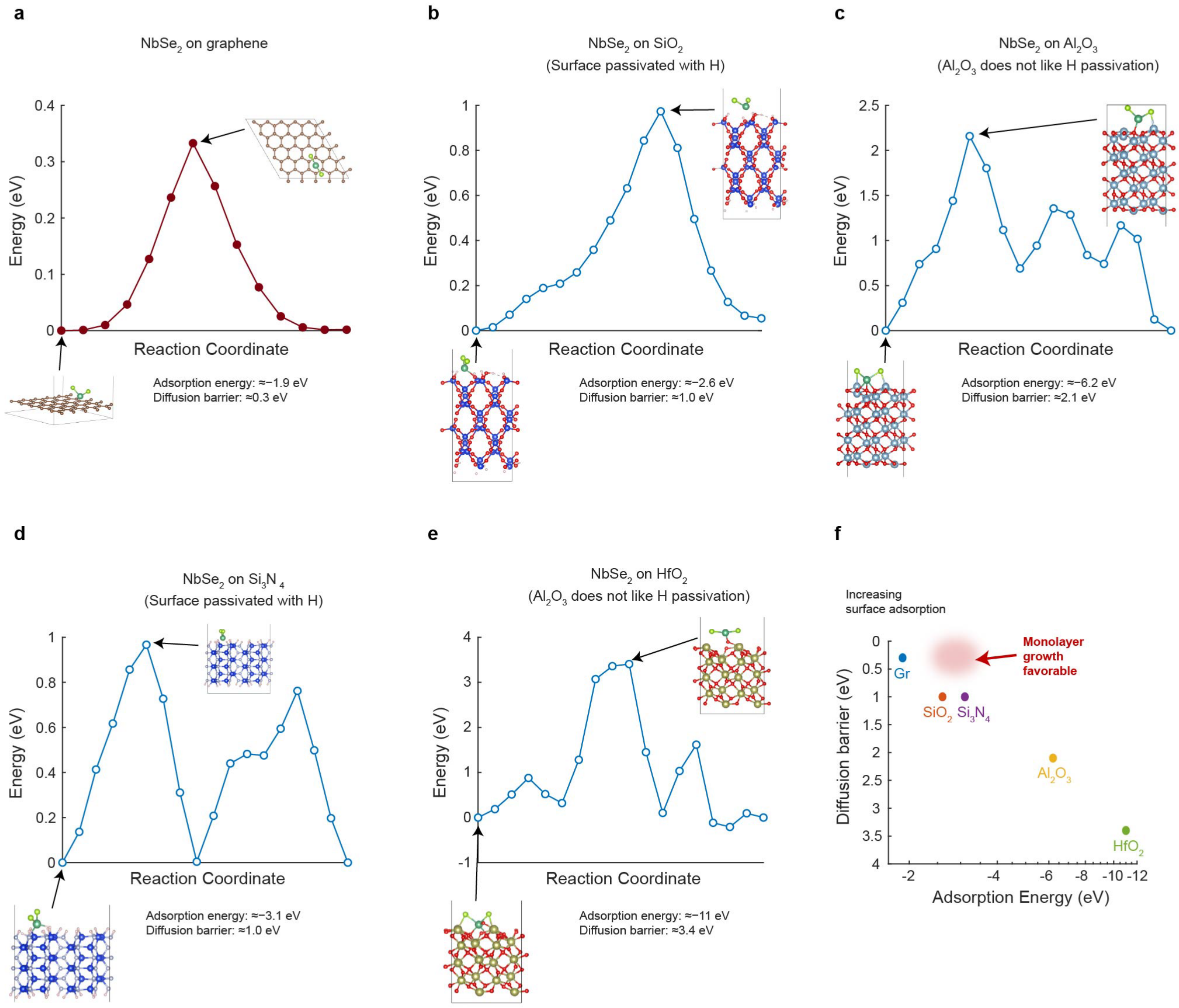

**Extended Data Fig. 3 | More comprehensive DFT calculations.** **a-e,** DFT-calculated energy along the diffusion path of $NbSe_2$ on **(a)** graphene, **(b)** $SiO_2$, **(c)** $Al_2O_3$, **(d)** $Si_3N_4$, and **(e)** $HfO_2$. **f**, Calculated diffusion energy vs. adsorption energy of $NbSe_2$ on various surfaces.

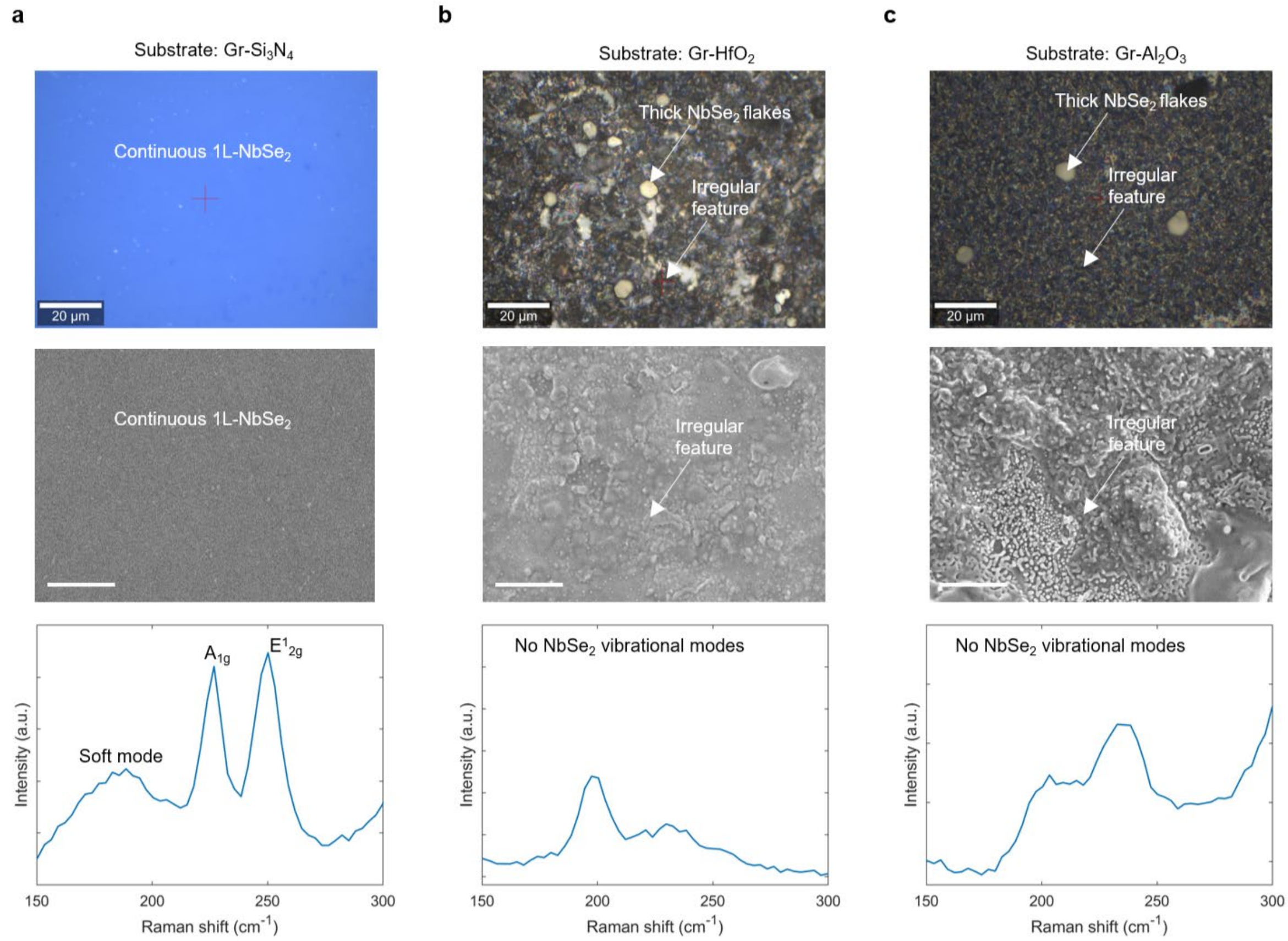

**Extended Data Fig. 4 | $NbSe_2$ growth test on different substrates. a-c,** Optical images, SEM images and Raman spectra showing the $NbSe_2$ growth result on (a) Graphene-$Si_3N_4$ substrate (b) Graphene-$HfO_2$ substrate and (c) Graphene-$Al_2O_3$ substrate. The continuous 1L-$NbSe_2$ is successfully grown at graphene-$Si_3N_4$ interface. But in the case of $HfO_2$ and $Al_2O_3$, only thick $NbSe_2$ flakes and irregular black features (no $NbSe_2$ Raman signal) were formed. The $Si_3N_4$ was grown by PECVD, and $HfO_2$ and $Al_2O_3$ were grown by ALD. The graphene films were then transferred on top of them. Scale bars for optical images, 20 µm. Scale bars for SEM images, 2 µm.

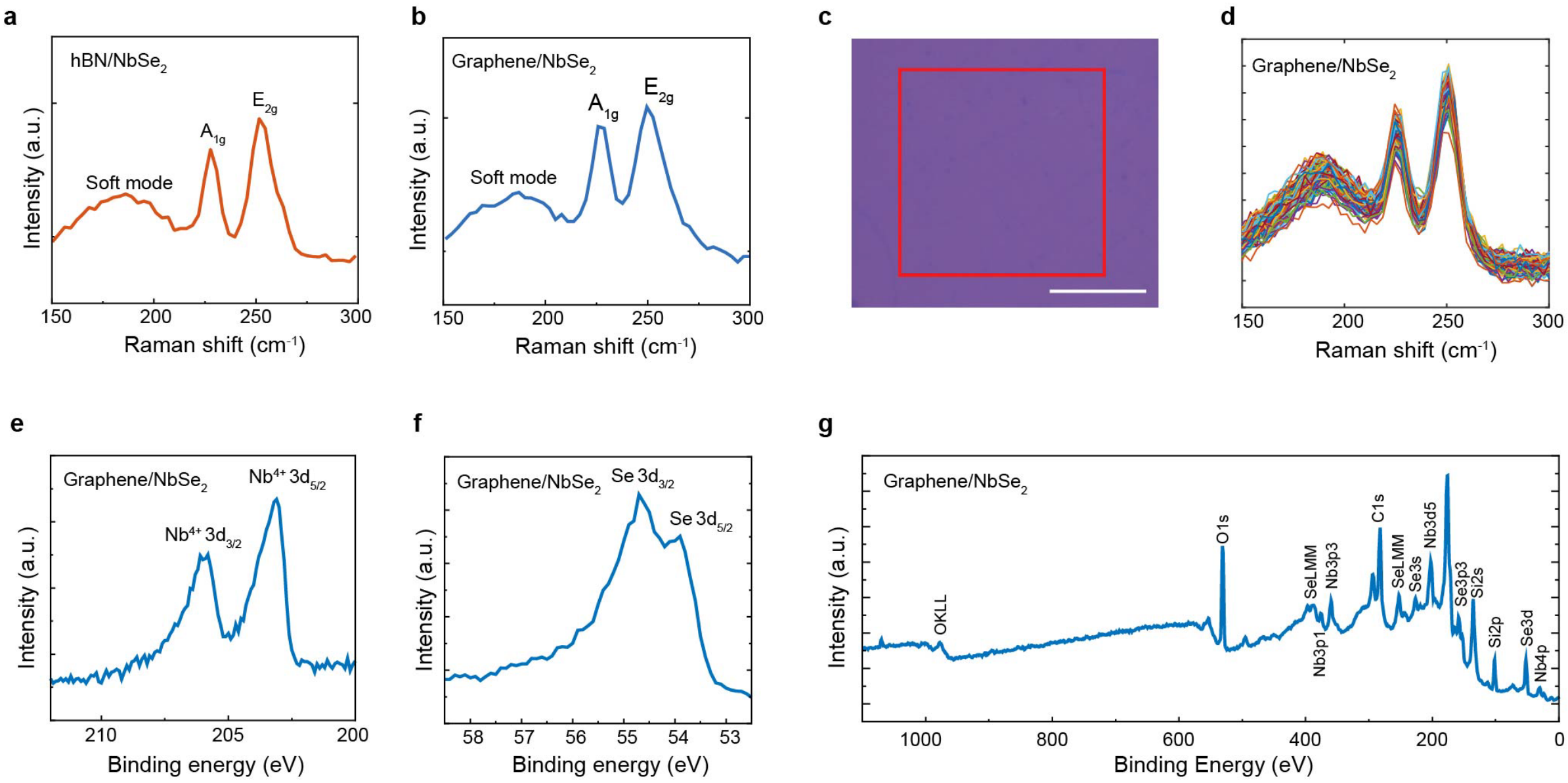

**Extended Data Fig. 5 | Raman and XPS characterization of as-grown $NbSe_2$. a,** Raman spectrum of as-grown hBN/$NbSe_2$. **b,** Raman spectrum of as-grown graphene/$NbSe_2$. **c,** Optical image showing the area of the graphene/$NbSe_2$ sample used for Raman mapping in Fig. 1i–j. Scale bar: 10 µm. **d,** Raman spectra collected from the mapped area shown in **c. e,** XPS spectrum of as-grown graphene/$NbSe_2$ showing Nb 3d peaks. Two peaks at 203.4 eV and 206.3 eV correspond to $Nb^{4+}$ $3d_{5/2}$ and $3d_{3/2}$ binding energies, consistent with literature values. **f,** XPS spectrum of as-grown graphene/$NbSe_2$ showing Se 3d peaks. Peaks at 53.2 eV and 54.2 eV correspond to Se $3d_{5/2}$ and $3d_{3/2}$ binding energies, also in good agreement with literature[16]. **g,** XPS Survey spectrum of the graphene/$NbSe_2$ sample.

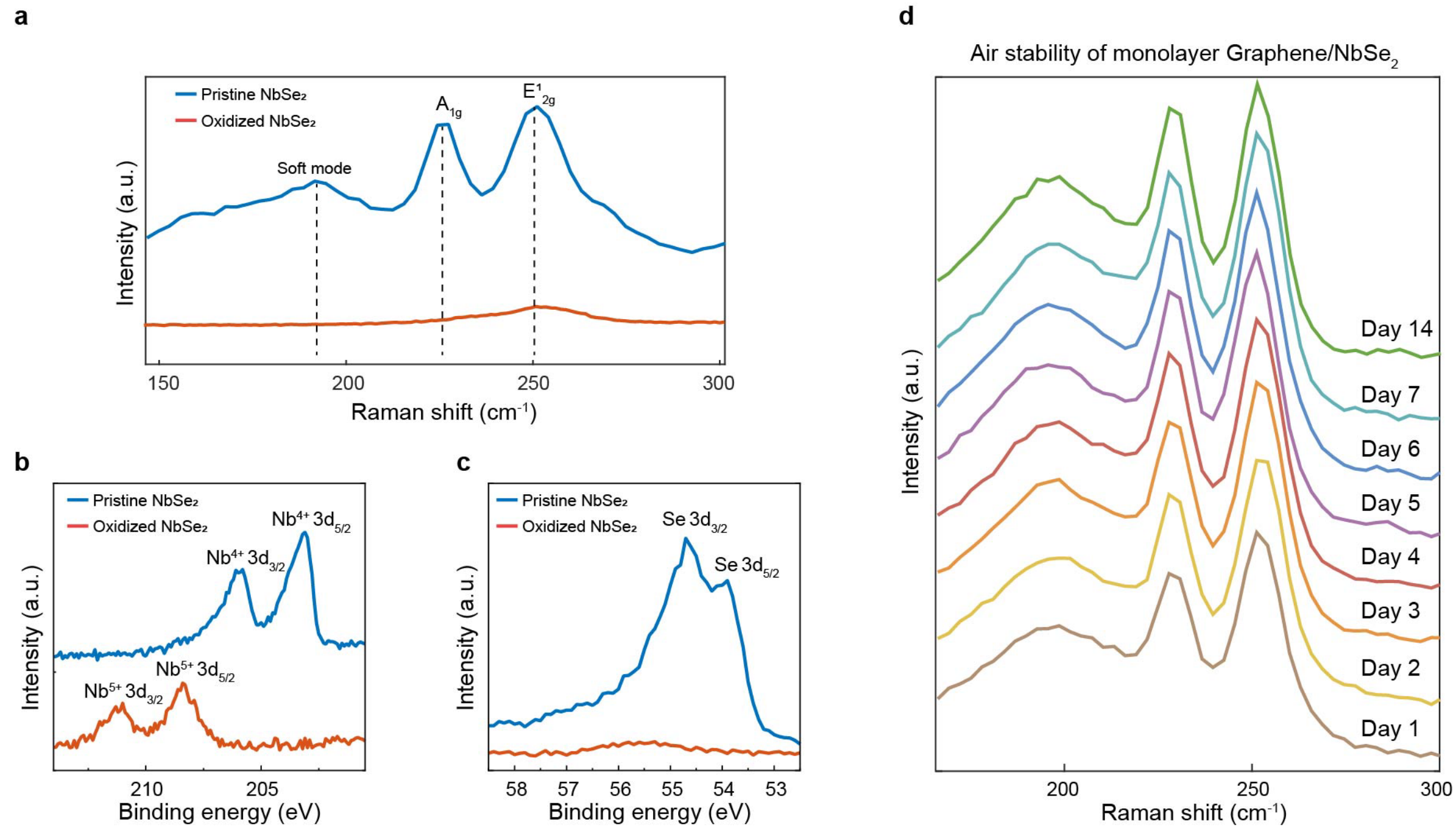

**Extended Data Fig. 6 | Air-stability characterization of 1L-$NbSe_2$ grown by encapsulation epitaxy. a,** Raman spectra comparing pristine $NbSe_2$ with intentionally oxidized $NbSe_2$. The disappearance of $NbSe_2$ vibrational modes upon oxidation highlights Raman spectroscopy as an effective tool for rapid air-stability assessment. **b-c,** X-ray photoelectron spectroscopy (XPS) of the same pristine and oxidized $NbSe_2$ samples shown in **b**. Upon oxidation, the Nb valence changes from $4^+$ to $5^+$, and the Se signal disappears, revealing that the oxidized $NbSe_2$ is composed of $Nb_2O_5$. Noted this $Nb_2O_5$ might be amorphous considering no Raman characteristic peak is detected. **d,** Raman spectra of a 1L-graphene/ $NbSe_2$ sample grown by encapsulation epitaxy. Minimal signs of oxidation are observed, despite continuous exposure to ambient environment throughout the period. This highlights the high air-stability of $NbSe_2$ in this work, which also serves as indirect evidence that $NbSe_2$ is grown underneath 2D template, as otherwise $NbSe_2$ will be immediately oxidized, which will diminish Raman signals (See Extended Data Fig. 2).

a

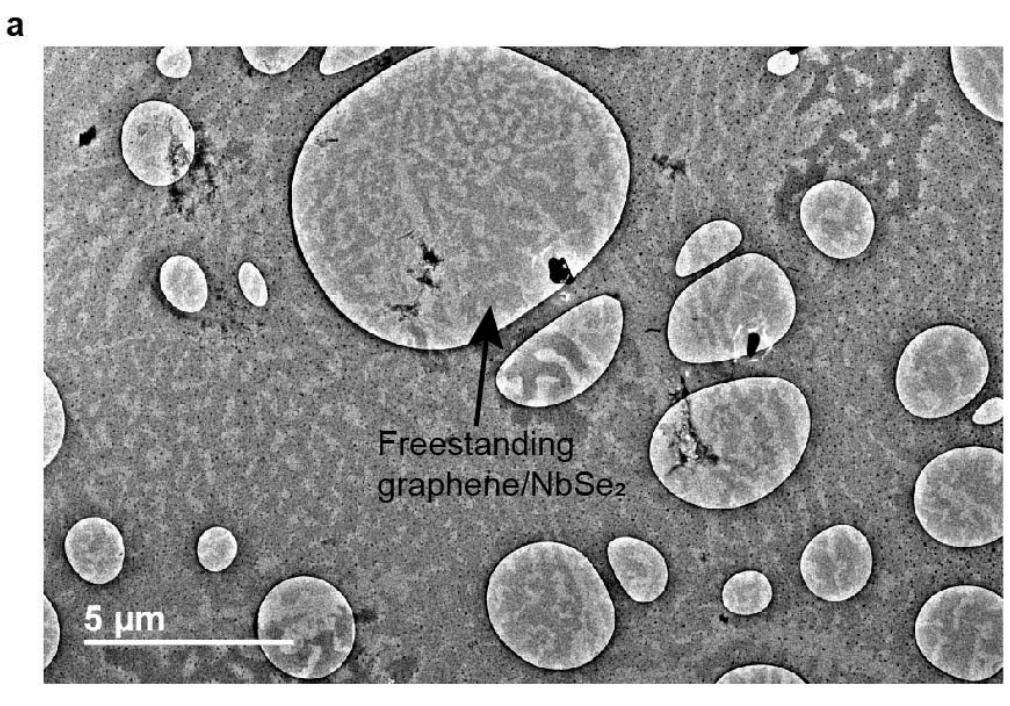

b

**Extended Data Fig. 7 | More SAED characterizations of graphene/$NbSe_2$.** a, Planar-view SEM image of graphene/$NbSe_2$ on TEM grid, showing one of the regions that SAED was taken. Graphene is a continuous monolayer while the $NbSe_2$ are partially grown flakes. The holes are free-standing areas. Scale bar, 5 µm. **b,** SAED measurements of graphene/$NbSe_2$ taken at random locations across a 200 × 200 $\mu m^2$ area. Scale bar, 5 $nm^{-1}$.

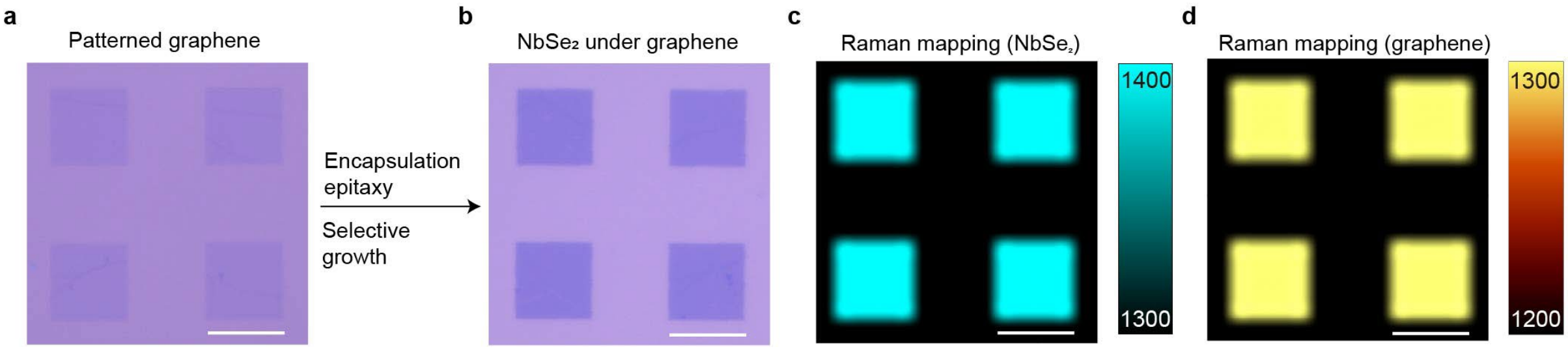

**Extended Data Fig. 8 | Selective growth of $NbSe_2$ utilizing patterned graphene. a,** Optical image of patterned graphene on $SiO_2$ substrate. **b,** Optical image of $NbSe_2$ grown underneath the patterned graphene shown in **a**. **c-d**, Raman mapping images of (**c**) $NbSe_2$ $E^1_{2g}$ peak and (**d**) graphene 2D peak. The Raman mapping was taken on the optical image shown in **b**. Scale bars, 10 μm.

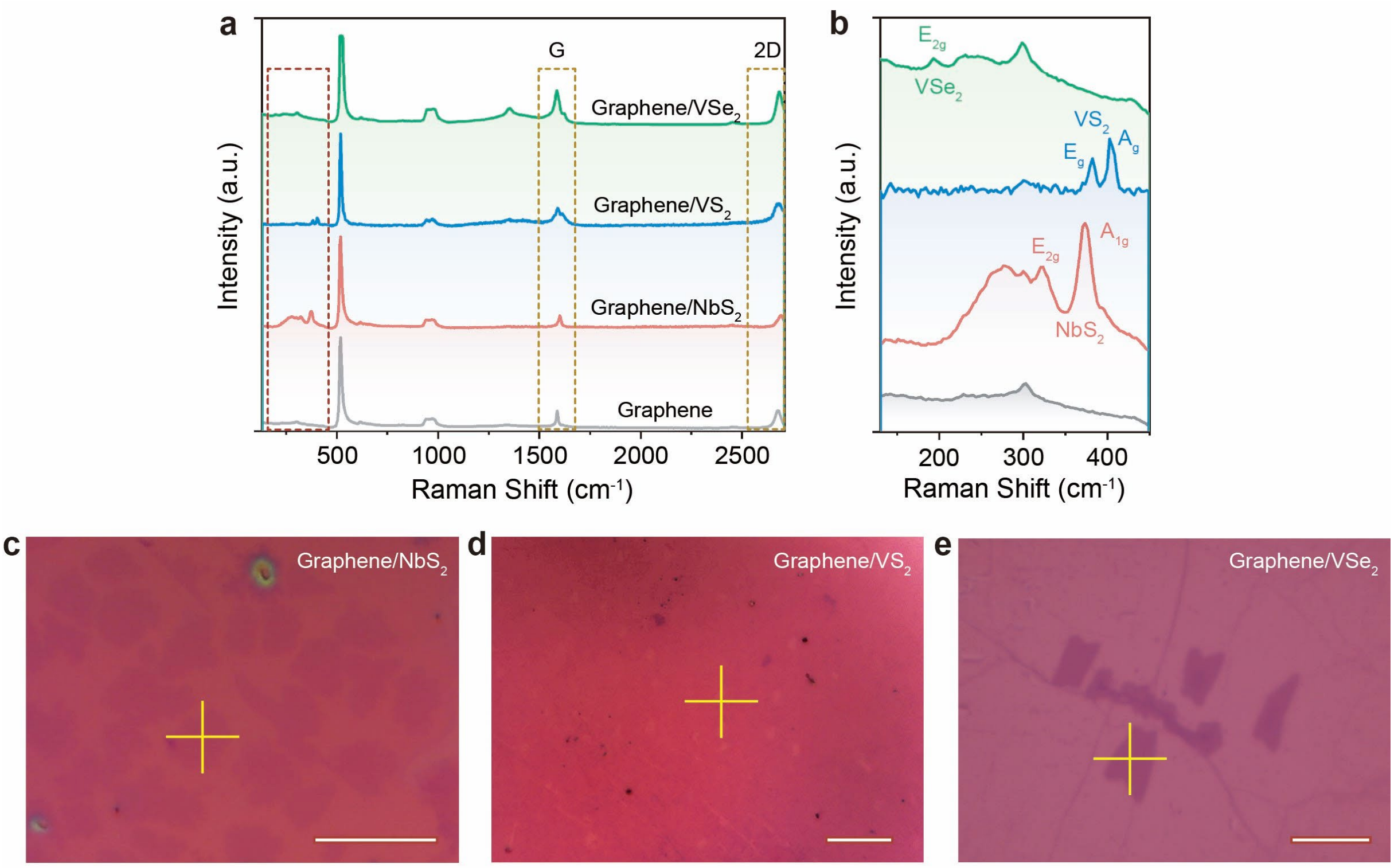

**Extended Data Fig. 9 | Growth of other high-melting-point TMDs by encapsulation epitaxy. a,** Raman spectra of as-grown graphene/$NbS_2$, graphene/$VS_2$, and graphene/$VSe_2$, with the spectrum of bare graphene included as a reference. Raman peaks located around 1590 and 2680 $cm^{-1}$ correspond to G and 2D vibration modes of graphene. Raman peaks at 300 and 520 $cm^{-1}$ arise from the silicon substrate. **b**, Zoomed-in view of panel **a** in the range 130–450 $cm^{-1}$ (the region within red dotted rectangle). The Raman peaks of graphene/$NbS_2$ are observed at 324 and 375 $cm^{-1}$, corresponding to the $E_{2g}$ and $A_{1g}$ modes of H-phase $NbS_2$[52]. The Raman peaks of graphene/$VS_2$ appear at 384 and 404 $cm^{-1}$, corresponding to the $E_g$ and $A_g$ modes of H-phase $VS_2$[53]. The Raman peak of graphene/$VSe_2$ appears at 193 $cm^{-1}$, corresponding to the $E_{2g}$ vibrational mode of H-phase $VSe_2$[54]. **c-e**, Optical images of as-grown graphene/$NbS_2$, graphene/$VS_2$ and graphene/$VSe_2$, respectively. The markers on the optical images indicate the measurement positions for the Raman spectra in panel a. Scale bars, 10 µm.

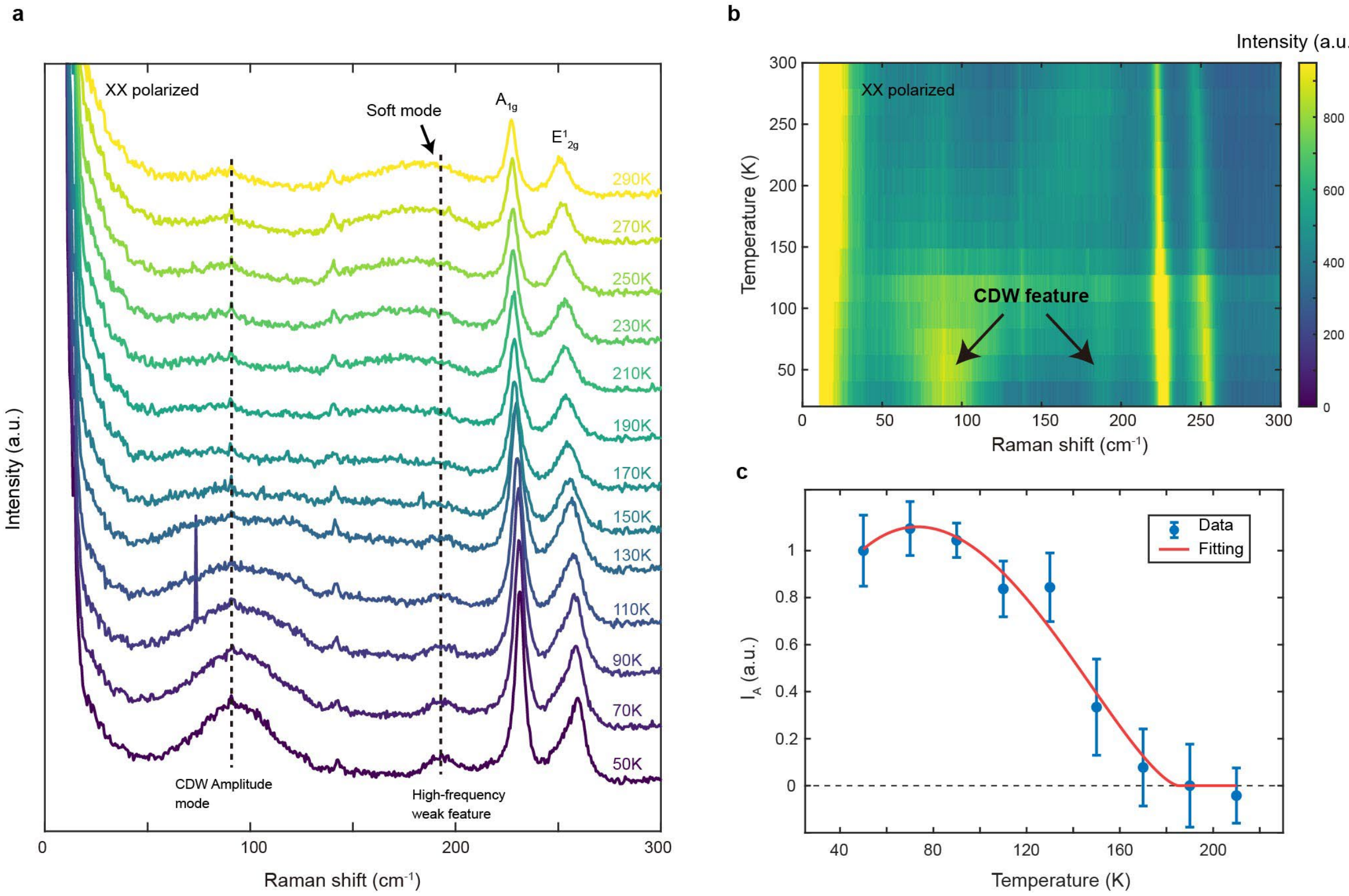

**Extended Data Fig. 10 | CDW measurement of hBN/$NbSe_2$. a,** Temperature dependent Raman measurement (range showing vibration modes of $NbSe_2$) with XX polarized laser on hBN/1L-$NbSe_2$ heterostructure synthesized by encapsulation epitaxy. The curves are vertically offset for clarity. **b,** Color plot of the temperature dependent Raman measurement shown in a. **c,** Temperature dependence of the CDW amplitude mode intensity ($I_A$) for 1L graphene/$NbSe_2$ heterostructure. The error bars are calculated based on the standard deviation of the Raman scattering intensity around the amplitude mode peak and the spectral width employed for integrating the mode. Solid lines are fits to mean field theory[3]. .